\documentclass[twocolumn, trackchanges]{aastex62}

\usepackage{url}
\usepackage{amsmath}
\usepackage{amssymb}
\usepackage{natbib}
\usepackage{multirow}
\usepackage{xcolor}
\usepackage{lineno}
\newcommand{\bvec}[1]{{\ensuremath{\boldsymbol{#1}}}}

\graphicspath{}

\received{}
\revised{}
\accepted{}
\submitjournal{}

\shorttitle{K2Rotation}
\shortauthors{Gordon et al.}
\begin{document}

\title{Stellar Rotation in the K2 Sample: Evidence for Modified Spindown}

\correspondingauthor{Tyler A. Gordon}
\email{tagordon@uw.edu}

\author[0000-0001-5253-1987]{Tyler A. Gordon}
\affiliation{Department of Astronomy, University of Washington, Box 351580, U.W., Seattle, WA 98195-1580, USA}

\author[0000-0002-0637-835X]{James R. A. Davenport}
\affiliation{Department of Astronomy, University of Washington, Box 351580, U.W., Seattle, WA 98195-1580, USA}

\author[0000-0003-4540-5661]{Ruth Angus}
\affiliation{Department of Astrophysics, American Museum of Natural History, 200 Central Park West, Manhattan, NY, USA}

\author[0000-0002-9328-5652]{Daniel Foreman-Mackey}
\affil{Center for Computational Astrophysics, Flatiron Institute, 162 5th Ave, New York, NY 10010}

\author[0000-0002-0802-9145]{Eric Agol}
\affiliation{Department of Astronomy, University of Washington, Box 351580, U.W., Seattle, WA 98195-1580, USA}

\author[0000-0001-6914-7797]{Kevin R. Covey}
\affiliation{Department of Physics \& Astronomy, Western Washington University, MS-9164, 516 High St., Bellingham, WA, 98225}

\author[0000-0001-7077-3664]{Marcel A.~Ag\"{u}eros}
\affiliation{Department of Astronomy, Columbia University, Manhattan, NY, USA}

\author[0000-0002-4365-7366]{David Kipping}
\affiliation{Department of Astronomy, Columbia University, Manhattan, NY, USA}



\begin{abstract}
    We analyze light curves of 284,834 unique K2 targets using a Gaussian process model with a quasi-periodic kernel function. By crossmatching K2 stars to observations from Gaia Data Release 2, we have identified 69,627 likely main-sequence stars. From these we select a subsample of 8,977 stars on the main-sequence with highly precise rotation period measurements. With this sample we recover the gap in the rotation period-color diagram first reported by \cite{McQuillan2013}. While the gap was tentatively detected in \cite{Reinhold2020}, this work represents the first robust detection of the gap in K2 data for field stars. This is significant because K2 observed along many lines of sight at wide angular separation, in contrast to Kepler's single line of sight. Together with recent results for rotation in open clusters, we interpret this gap as evidence for a departure from the $t^{-1/2}$ Skumanich spin down law, rather than an indication of a bimodal star formation history. We provide maximum likelihood estimates and uncertainties for all parameters of the quasi-periodic light curve model for each of the 284,834 stars in our sample. 
\end{abstract}

%\keywords{}

%\tableofcontents
\section{introduction}\label{sec:intro}

    Stellar rotation is a key physical property for understanding individual stars as well as stellar populations. Rotation drives the stellar dynamo which produces surface magnetic fields. These magnetic fields in turn give rise to stellar activity (e.g. starspots and flares). The rotation period ($P_\mathrm{rot}$) of a star is tied to its age through magnetic braking, which slows the star's rotation over time \citep{Durney1972, Skumanich1972}. Age is a fundamental stellar parameter, but is difficult to determine from the position of a star on a color-magnitude diagram, especially for stars on the main-sequence. Any information we can extract about the age of a star from its rotation period is therefore very valuable. This is the subject of gyrochronology, which seeks to measure stellar ages by observing the star's rate of rotation \citep{Barnes2003}. 
    
    The Kepler mission \citep{borucki2010} revolutionized the study of stellar rotation by producing high-precision light curves for hundreds of thousands of stars, from which over 34,000 rotation periods have been inferred \citep{nielsen2013, McQuillan2014}. The distribution of rotation periods measured by \citet{McQuillan2013} showed an unexpected bimodality in the field M dwarfs, which was found to extend to the K dwarfs by \citet{McQuillan2014}. This bimodality was recovered in the G dwarfs by \cite{Davenport2018}, who used Gaia astrometry to limit their analysis to main-sequence stars with well-determined Gaia photometric solutions, removing contamination by subgiants.
    
    Several explanations have been put forward to explain this bimodal period distribution. \cite{Davenport2018} and \cite{McQuillan2013, McQuillan2014} suggest that the bimodality may be the result of a bimodal star formation history, with a recent burst of star formation accounting for the fast-rotating branch of the bimodality and an older population of stars forming the slow-rotating branch. \cite{Reinhold2019} propose that the gap between modalities may represent a minimum in detectability of rotation periods due to the transition from spot-dominated to faculae-dominated stellar activity. 
    
    A third possibility, to be discussed in more detail in the Section \ref{sec:discussion}, is that the gap results directly from the spin evolution of G, K and M dwarfs. An epoch of stalled spindown followed by a period of rapid angular momentum loss before the resumption of Skumanich spindown may be able to explain such a feature. 
    To date, rotation periods from open clusters have provided the most compelling evidence that modified spin-evolution is indeed the cause of the gap. These fixed-age populations have shown that rotation periods for low-mass stars break from the expected Skumanich spindown model, such as the cluster of rotation periods at $P_\mathrm{rot}\approx10$ days found in the 1 Gyr old cluster NGC 6811 by \cite{Meibom2011}. Similar deviations from the traditional Skumanich spindown profile have been seen in e.g. Praesepe at 650 Myr \citep{Douglas2017}
    and NGC 752 at 1.3 Gyr \citep{Agueros2018}. 
    \citet{Curtis2019} note in their analysis of NGC 6811 that the stall in spin-down appears to be mass and age dependent. Further, \citet{Curtis2020} show compelling evidence that this deviation from Skumanich spin-down indeed corresponds to the gap in field rotation periods, with individual cluster sequences ``crossing'' the rotation period gap.
    
    This scenario may be explained in terms of time-variable and mass-dependent rotational coupling between the core and envelope of the star \citep{Spada2020}. Magnetic braking slows the rotation of the convective envelopes of stars. However, if the core and envelope are only weakly coupled, the stellar core may continue to spin rapidly even as the envelope slows down. A decoupled core and envelope with reduced angular momentum exchange is expected for young stars \citep{Endal1981, MacGregor1991, Bouvier2008, Denissenkov2010, Gallet2013, Lanzafame2015, Somers2016}. After a period of time, which appears to depend on stellar mass, the core and the envelope begin exchanging angular momentum. When angular momentum transport is happening efficiently, the core’s angular momentum transferred to the envelope would offset magnetic braking, allowing the envelope to maintain a constant rotation period. After the core and envelope have coupled and the star rotates as a solid body, the star would resume spinning down. This process would result in a departure from the Skumanich spindown law \citep{Skumanich1972}, which prescribes a smooth spindown over time following the relation 
    \begin{equation}
        P_\mathrm{rot}\propto t^{-1/2},
    \end{equation}
    where $P_\mathrm{rot}$ is the stellar rotation period as a function of the age of a star, $t$. This scenario may explain the convergence of cluster sequences below the gap. Explaining the under-density within the gap requires us to posit an additional stage in stellar spin evolution consisting of a period of accelerated spindown immediately after the epoch of stalled spindown and before the resumption of Skumanich spindown \citep{Curtis2020}. This accelerated spindown is not predicted by the coupling scenario presented here, and its explanation will likely require further theoretical work.
    
    The Kepler data alone gives us a limited ability to explore these various hypotheses due to its single pointing, which admits the possibility that the bimodality is unique to the Kepler field. In contrast, K2 observed the sky in 18 separate campaigns, each having a different line of sight (save for a few overlapping campaigns). As \cite{vanSaders2019} note, if the Kepler line of sight happened to point directly through a late burst of star formation, thereby accounting for the fast rotating branch of the bimodal period distribution, we would not expect this feature to be visible in all 18 K2 campaigns.

    In this work we measure and report probabilistic constraints on periodic signals for 284,834 K2 stars from all 18 campaigns and analyze a subset of 8,943 highly accurate rotation periods. For those stars appearing in multiple campaigns we run our analysis separately for each light curve. We use a modification of the Gaussian Process regression method described in \cite{Angus2018} to measure periodic signals. We find that the bimodality is visible in all K2 campaigns, lending support to the idea that the feature is related to stellar physics rather than being a product of the star formation history within the Kepler field.

\section{Measuring Rotation Periods}\label{sec:measuring}

    We begin by describing the model that we use to infer probabilistic rotation periods from the EVEREST light curves \citep{Luger2018}. Stellar magnetic activity induces starspots and faculae on the star's surface. As the star rotates, these features are carried into and out of view, introducing periodicity into the light curve. If the starspots and faculae were static over time, we would observe a perfect periodicity with the star returning to the same luminosity once every period. However, starspots and faculae are not static, but rather evolve over time, emerging, changing shape, and disappearing as the star's rotation brings them into and out of view. As a result, the light curves do not display perfectly periodic variations, but rather a quasi-periodic variability with the shape and amplitude of the variability changing from period to period. This means that inferring rotation periods using straightforward sinusoidal variability models does not give good results. Instead, non-inference based methods such as autocorrelation functions \citep{McQuillan2013}, or Lomb-Scargle periodigrams \citep{Reinhold2013} can be employed. An alternative to these non-inference methods is to use a stochastic variability model such as a Gaussian process \citep{Angus2018}. In this work we use the autocorrelation function to derive a multi-modal ``prior'' over the period. We then use a Markov chain Monte Carlo (MCMC) method to estimate prosteriors for the parameters of the quasi-periodic Gaussian Process (GP) variability model defined in section \ref{sec:gp_analysis}. The use of quotation marks around the word ``prior'' references the fact that this is not technically a Bayesian prior, because it does not strictly depend on our prior beliefs about the period distribution. We explain this further and describe the effect that this has on our analysis in the next section. 
    
    \subsection{Autocorrelation Function Analysis}
    
    We use long cadence K2 EVEREST light curves from \cite{Luger2018} as the starting point for our analysis. We use the co-trending basis vector corrected flux (keyword FCOR in the EVEREST FITS files) which removes systematic trends from the raw light curves. To further remove long-term trends we subtract a third order polynomial from each light curve before computing the autocorrelation function (ACF). This has the effect of flattening the overall decay of the ACF at short time lags, improving our ability to detect the rotation period from the ACF peak. We note that the third order polynomial may overfit and remove some rotation signals for the slowest rotators in our sample. This primarily affects rotation periods longer than about 25 days, which does not interfere with our detection or analysis of the period gap in section \ref{sec:gap}. In our initial experiments we found that a second order polynomial did not sufficiently flatten the decay of the ACF and higher order polynomials were too likely to overfit the rotation signal. We remove outliers from our light curves by masking all flux observations greater than 3-sigma from a running median with a kernel width of 5 K2 long cadences.
    
    After pre-processing the light curve as described above, we compute the ACF for each light curve using the implementation provided in \texttt{exoplanet}, which wraps the \texttt{astropy} ACF function \citep{exoplanet:astropy18, exoplanet:exoplanet}. We smooth the ACF with a Gaussian filter with a kernel width of 0.5 days. We then use the smoothed ACF to construct a unique multi-modal period ``prior'', which we find aids in convergence during the MCMC step. We place ``prior'' in quotes because, strictly speaking, a prior should only reflect our prior beliefs about the period distribution rather than depending on the data itself. The distribution we derive here is not technically a prior although we use it as such. Since we are building our ``prior'' from the same data that we use to fit the GP, we risk underestimating the uncertainty on the period. Since our analysis focuses on point estimates of the period rather than the full posterior, we elected to accept this risk in exchange for the benefit of recovering more rotation periods. 
    
    Our prior is a Gaussian mixture with $3N$ components, where $N$ is given by 
    \begin{equation}
        N = \begin{cases}
            N_\mathrm{peaks}(\mathrm{p > 0.01}) & N_\mathrm{peaks}(\mathrm{p > 0.01}) < 10 \\
            10 & N_\mathrm{peaks}(\mathrm{p > 0.01}) \geq 10
        \end{cases}
    \end{equation}
    where $N_\mathrm{peaks}(p)$ is an integer corresponding to the number of peaks in the ACF with topographical prominence greater than $p$. The topographical prominence is computed with respect to the adjacent ACF minima, by \texttt{scipy}'s \texttt{signal} library \citep{scipy}. We take each of these peaks to represent a candidate period, recognizing that for a well-defined periodic signal there will be multiple peaks corresponding to the same period. 
    
    The factor of $3$ in $3N$ arises from our inclusion of candidate periods at $\tau_i/2$ and $2\tau_i$ where 
    $\tau_i$ is the lag of the $i^\mathrm{th}$ peak, so 
    that for each peak we have $3$ candidate periods.
    The weight of the component of the Gaussian mixture prior corresponding each peak is given by 
    \begin{equation}
        w_i = h_i\sqrt{p_i} 
    \end{equation}
    where $h_i$ is the height of the peak, and is the same for the candidate periods at $\tau_i/2$ and $2\tau_i$ as for the candidate period at $\tau_i$ itself. The standard deviation of each Gaussian component, $\sigma_i$, is given by the width of the peak at half of the peak height. This means that the width of the Gaussian component of the prior is wider by a factor of approximately 2.35 than the standard deviation of the Gaussian equivalent to the ACF peak. The standard deviation associated with the candidate period at $\tau_i$ is also used for the candidate periods at $2\tau_i$ and $\tau_i/2$. In the case that no peaks are detected in the ACF, we adopt a uniform prior over the range $P=(0, \Delta T/2)$ where $\Delta T$ is the total duration of the light curve.
    
\begin{figure*}
        \centering
        \includegraphics[width=\textwidth]{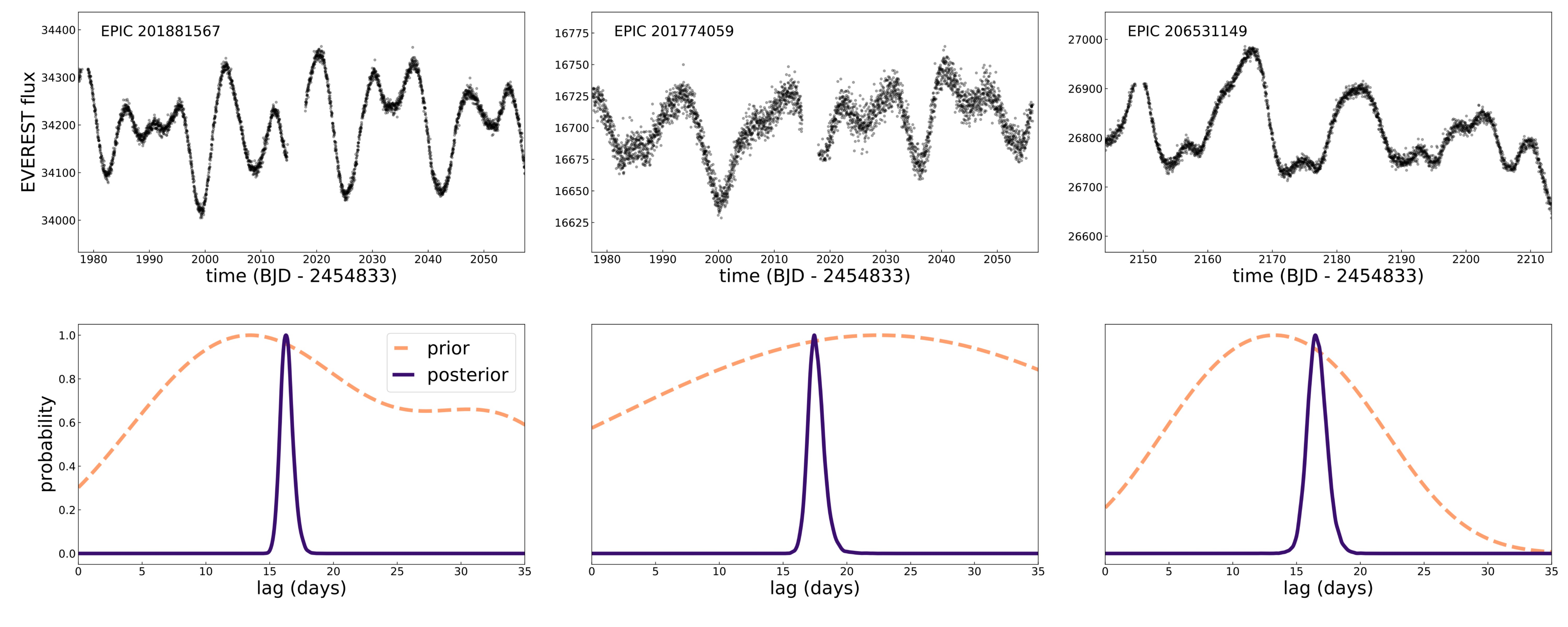}
        \includegraphics[width=\textwidth]{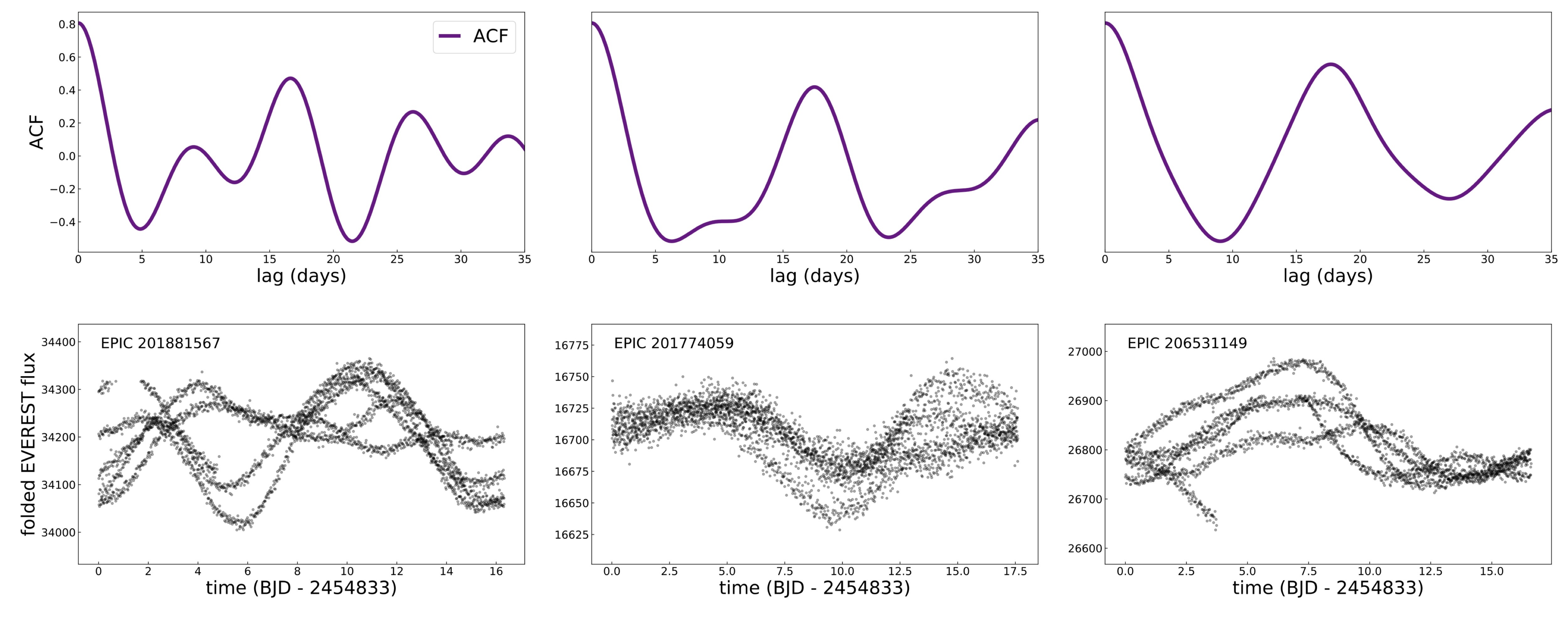}
        \caption{Sample output from our period detection 
        procedure for three K2 stars with well-determined 
        rotation periods. The top row of panels show the cotrending basis 
        vector-detrended EVEREST flux. The second row shows the 
        period prior and MCMC-estimated posterior. The third row shows the autocorrelation function, and the bottom row shows the light curve folded on the mean of the posterior for the period.} For visibility of the prior, both the prior and posterior are normalized such that their maximum probability is 1.
        \label{fig:example_detections}
    \end{figure*}

    The choices we made in computing $w_i$ and $\sigma_i$ are motivated by the logic that a higher peak should be given more weight in the mixture than a lower peak, and that the component in the mixture should have a smaller standard deviation if the peak in the ACF is sharper, reflecting the smaller uncertainty on the corresponding period. We also aim to construct a prior that is informative, but that doesn't prohibit the MCMC from exploring periods not identified as period candidates by our algorithm. As can be seen in Figures \ref{fig:example_detections}, the period priors we construct tend to have wide regions of high probability and therefore limited influence over our point estimates of the rotation period. 

    %\begin{figure*}
    %    \includegraphics[width=\hsize]{figures/example%_nondetections.pdf}
    %    \caption{Sample output from our period %detection 
    %    procedure for three K2 stars without %detectable 
    %    rotation.}
    %    \label{fig:acf_examples_nondetection}
    %\end{figure*}
    
    \subsection{Gaussian Processes and MCMC Analysis}
    \label{sec:gp_analysis}
    
    A Gaussian process can be thought of as a distribution from which we may draw random functions with a given covariance structure. They are commonly used in astrophysics to model stochastic variability in light curves (see \citealt{Dawson2014}, \citealt{Barclay2015}, and \citealt{Chakrabarty2019} for examples from studies of transiting exoplanets and \cite{MacLeod2010} for an example in which a GP is used to model AGN variability). A GP can be split into two components: a kernel function $k(\tau)$ which describes the covariance of the functions in the distribution, and a mean function $\mu(t)$. The kernel function defines the covariance matrix of a multi-dimensional Gaussian distribution by specifying the covariance between every pair of flux measurements. The covariance matrix is given by
    \begin{equation}
        K_{i, j} = k(\tau_{i, j}).
    \end{equation}
    where $\tau_{i,j} = |t_i-t_j|$ is the absolute value of the separation between times $t_i$ and $t_j$ and $K_{i, j}$ is the $i^\mathrm{th}, j^\mathrm{th}$ entry of the covariance matrix $K$.  
    
    The log of the likelihood function of the GP is given by 
    \begin{equation}
	 \label{eqn:simple_logL}
		\ln\ \mathcal{L} = 
		-\frac{1}{2}(\bvec{y}-\bvec{\mu})^\mathrm{T} K^{-1}(\bvec{y}-\bvec{\mu})  
		-\frac{1}{2}\ln\ \mathrm{det}(K) - \frac{N}{2}\ln(2\pi)
	\end{equation}
	where $\bvec{y}$ is a vector of observations and $\bvec{\mu}$ is a mean vector with the same length as $\bvec{y}$. Both the kernel and mean are parameterized by a set of hyperparameters. GP regression is the process of finding the hyperparameters that maximize the GP's likelihood with respect to a set of observations. For a more detailed primer on Gaussian processes in astronomy, see \cite{Foreman-Mackey2017}, or, for a more complete resource on Gaussian Processes across fields, we refer the reader to \citet{Rasmussen2006}. 
    
    To construct a GP stellar rotation model we take the mean function of the GP to be constant and allow the kernel function to model the correlated variability introduced into the star's light curve by spots and faculae as they rotate in and out of view and evolve over time. Our GP model has three terms: two quasi-periodic terms to capture the rotationally induced variability and one that is aperiodic to capture any leftover variability originating from other astrophysical sources or instrumental effects. The power spectrum of each term is given by 
    \begin{equation}
        S_i(\omega) = \sqrt{\frac{2}{\pi}}\frac{S_i\omega_i^4}{(\omega^2 - \omega_i^2)^2 + 2\omega^2\omega_i^2}.
    \end{equation}
    For the periodic terms we follow \cite{Foreman-Mackey2017} in setting
    \begin{eqnarray}
        \label{eqn:kernel}
        \nonumber Q_1 &=& 1/2 + Q + \Delta Q \\ \nonumber \\ \nonumber
        Q_2 &=& 1/2 + Q \\ \nonumber \\ \nonumber
        \omega_1 &=& \frac{4\pi Q_1}{P\sqrt{4Q_1^2 - 1}} \\ \nonumber \\
        \omega_2 &=& \frac{8\pi Q_1}{P\sqrt{4Q_1^2 - 1}} \\ \nonumber \\ \nonumber
        S_1 &=& \frac{\sigma^2}{(1 + f)\omega_1Q_1} \\ \nonumber \\ \nonumber
        S_2 &=& \frac{f\sigma^2}{(1 + f)\omega_2Q_2}  \nonumber
    \end{eqnarray}
    where $Q$ is the quality factor, $\Delta Q$ specifies the offset in the quality factor between the two oscillators, $P$ is the period of the oscillator, $\sigma^2$ is the variance of the oscillation, and $f$ specifies the fractional contribution of the oscillator at the half period $P/2$ compared to the oscillator at the full period. For the aperiodic term we set:
    \begin{eqnarray}
            Q_3 = 1/\sqrt{2}
    \end{eqnarray}
    while $\omega_3$ and $S_3$ are free parameters. Setting $Q_3=1/\sqrt{2}$ means that the third oscillator is critically damped and will not display periodic oscillations. For this term the power spectrum simplifies to 
    \begin{equation}
        S_3(\omega) = \sqrt{\frac{2}{\pi}}\frac{S_3}{(\omega/\omega_3)^4+1}.
    \end{equation}
    The full variability model including both the quasi-periodic and aperiodic terms has the power spectrum 
    \begin{equation}
        S(\omega) = \sum_i^3 S_i(\omega).
    \end{equation}
    
 To compute the GP model we use the \texttt{celerite} GP method \citep{Foreman-Mackey2017} as implemented in \texttt{exoplanet} \citep{exoplanet:exoplanet}. We maximize the GP likelihood with respect to the EVEREST co-trending basis vector detrended flux for the parameters $\{P, Q, \Delta Q, A, f, S_3, \omega_3\}$. We then use the maximum likelihood solution as a starting point for our MCMC analysis. We use uninformative priors for all GP hyperparameters except the period, for which we use the multi-modal Gaussian mixture prior described previously. We use the NUTS sampler provided by PyMC3 \cite{exoplanet:pymc3} to run 1000 tuning samples followed by 500 production samples on each of 28 cores for a total of 28,000 tuning and 14,000 production samples. We have found that a relatively large number of tuning samples is helpful for achieving convergence when using a multi-modal period prior, in order to allow the sampler to fully explore the multi-modal likelihood space. 
    
\begin{figure*}
        \centering
        \includegraphics[width=\hsize]{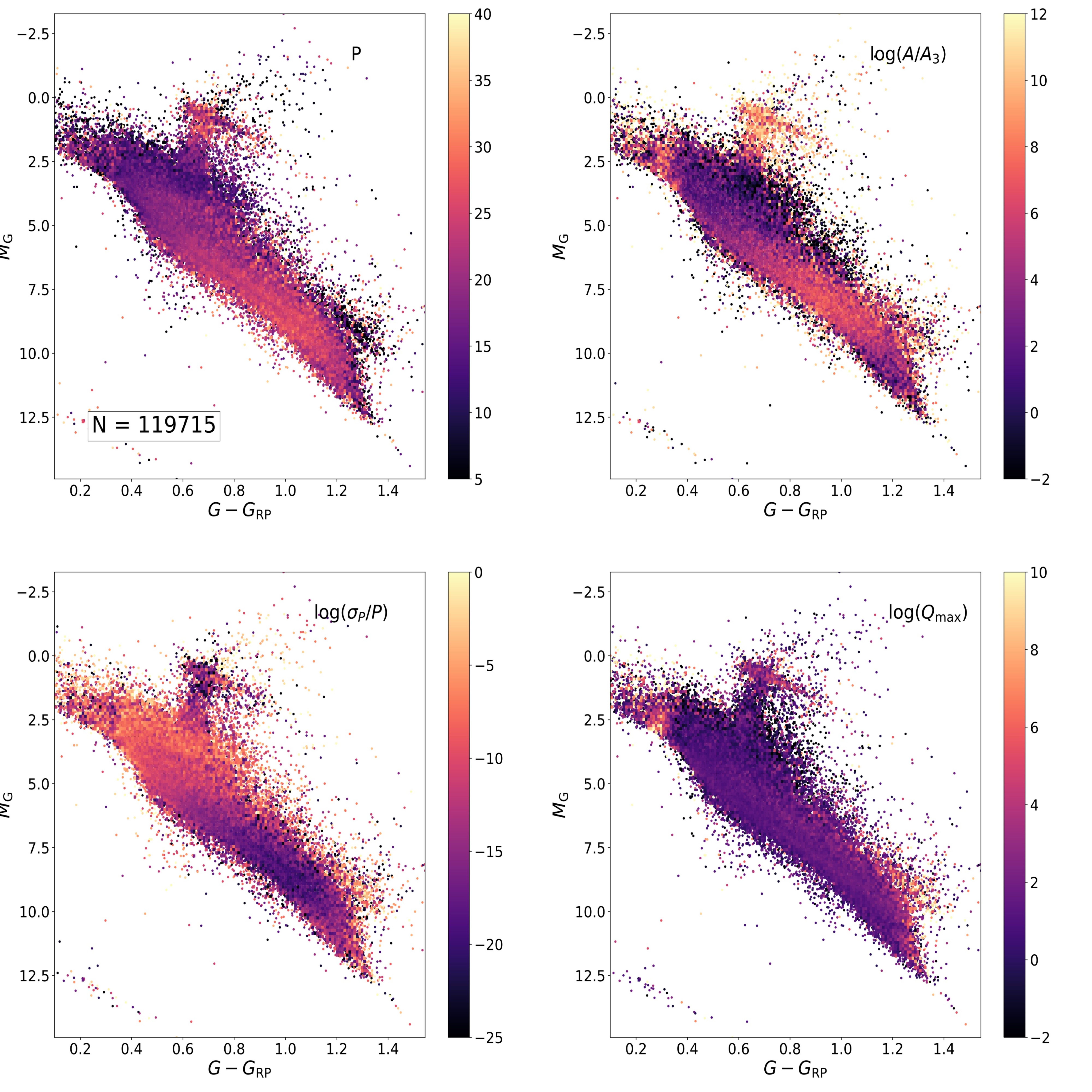}
		\caption{Selected hyperparameters plotted over the Gaia color-magnitude diagram. For each plot the color in a bin indicates the mean of the quantity given in the upper right-hand corner of the plot and the scattered points are colored by that quantity in regions where the density of stars is low. \textbf{Upper left:} Rotation period. \textbf{Upper right:} Log of the ratio between the periodic variance, $A$, and the variance of the aperiodic component, $A_3$. \textbf{Lower left:} Log of the fractional uncertainty for the inferred period. \textbf{Lower right:} Log of the mean quality factor, $Q_\mathrm{max}$, with larger $Q_\mathrm{max}$ indicating stronger periodicity.}
		\label{fig:cmds}
	\end{figure*}

In Figure \ref{fig:cmds} we show the variation in the binned mean of the period $P$, maximum quality factor $Q_\mathrm{max} = \mathrm{max}(Q_1, Q_2)$, the logarithm of the ratio between the periodic and aperiodic components of the model, and the logarithm of the fractional uncertainty in rotation period.
    
	\subsection{Selecting Main Sequence Stars}
    \label{sec:main_sequence}
    
    We begin by making selections based on the quality of the Gaia DR2 photometric solutions \citep{Gaia2018}. We require that the following conditions be met: 
    \begin{itemize}
        \item $\sigma(G)/G < 0.01$
        \item $\sigma(G_\mathrm{RP})/G_\mathrm{RP} < 0.01$.
    \end{itemize}
    where $G$ and $G_\mathrm{RP}$ refer to the passbands used in Gaia DR1 and DR2. 
    
\begin{figure}
        \centering
        \includegraphics[width=1.05\hsize]{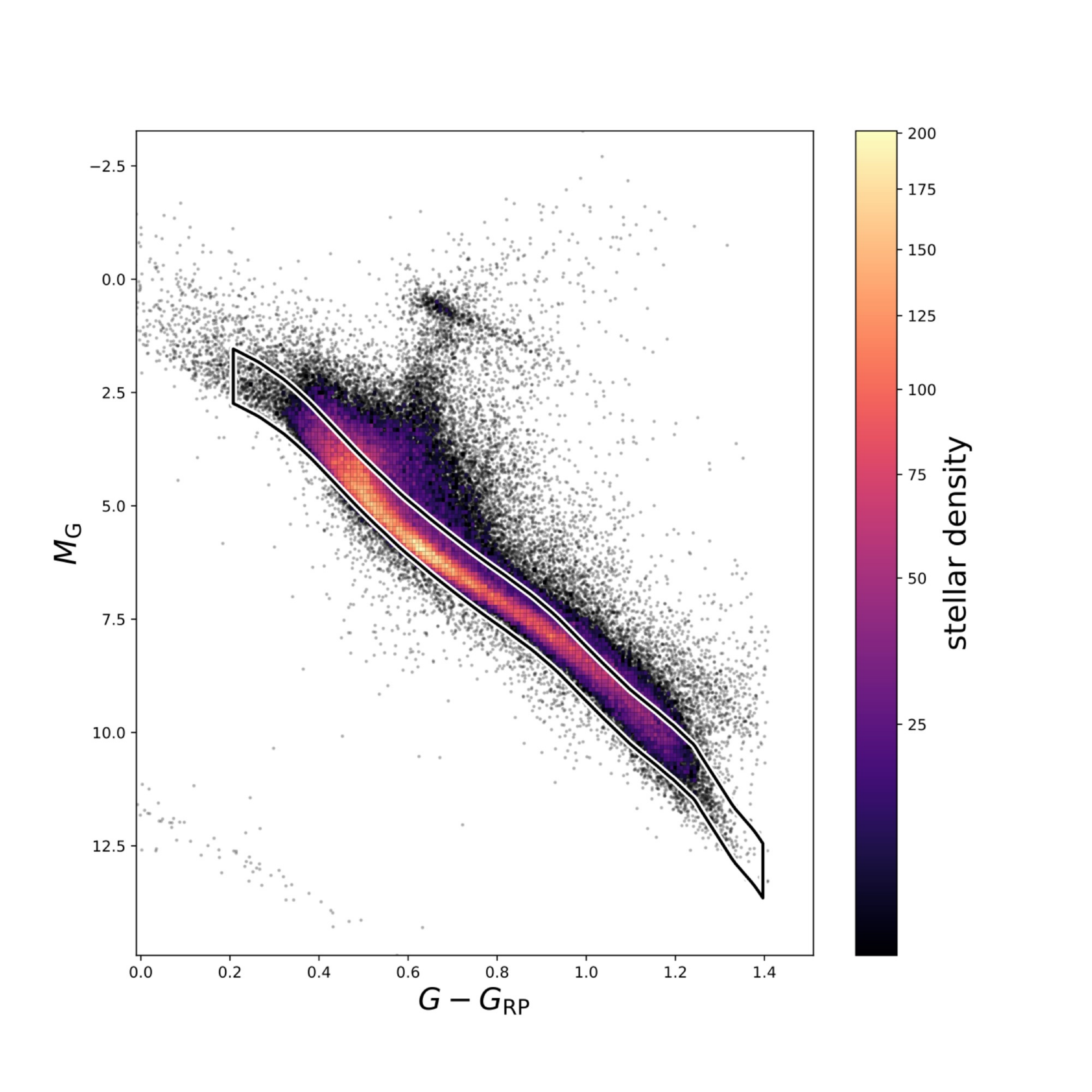}
		\caption{K2 stars on the Gaia color-magnitude diagram. The boxed region shows the area selected as the main sequence in order to exclude, e.g., evolved stars and unresolved binaries from our final sample. The main sequence is defined by a MIST isochrone with an age of $10^9$ years and $[\mathrm{Fe}/\mathrm{H}] = +0.5$. We identify 123,079 stars belonging to the main sequence, of which 8,943 meet the requirements for our final sample.}
		\label{fig:cmd}
	\end{figure}

    In order to reduce contamination from giants, subgiants and unresolved binary stars, we require that stars in our sample be on or near the main sequence, as defined by a MIST isochrone \citep{Choi2016, Dotter2016, Paxton2011, Paxton2013, Paxton2015} with an age of 200 Myr and a metallicity of $[\mathrm{Fe}/\mathrm{H}] = +0.25$ to identify the nominal main sequence, and we select stars within 0.3 mag below and 0.9 mag above the isochrone, as shown in Figure \ref{fig:cmd}. This wide slice of magnitude space allows us to encompass different ages and metallicities while reducing contamination from the giant and subgiant branches. The cost of selecting such a wide slice is that we likely incorporate a significant number of unresolved binaries into our final sample, which will add some amount of contamination. We find this acceptable since we don't expect this contamination to have a systematic influence on the overall shape of the period-color diagram.
    
    It should be emphasized that we made no attempt to choose an isochrone that represents the actual main-sequence for stars in our sample, which would be infeasible because the K2 sample contains stars with a wide range of ages and metallicities. The choice of $[\mathrm{Fe}/\mathrm{H}] = +0.25$ was made on the basis that this isochrone does an adequate job of matching the trend of the main-sequence. We have found that the exact choice of age and metallicity does not have a significant impact on our results, so long as the isochrone and the width of the box in $M_G$ selects a sufficient number of stars for our edge finding algorithm to perform well. Our final sample consists of 8,943 stars near the main sequence, passing our Gaia photometry cuts and possessing well-determined periodicity. 
    
    \subsection{Vetting Rotation Periods}
    \label{sec:vetting}
    
\begin{figure}
        \plotone{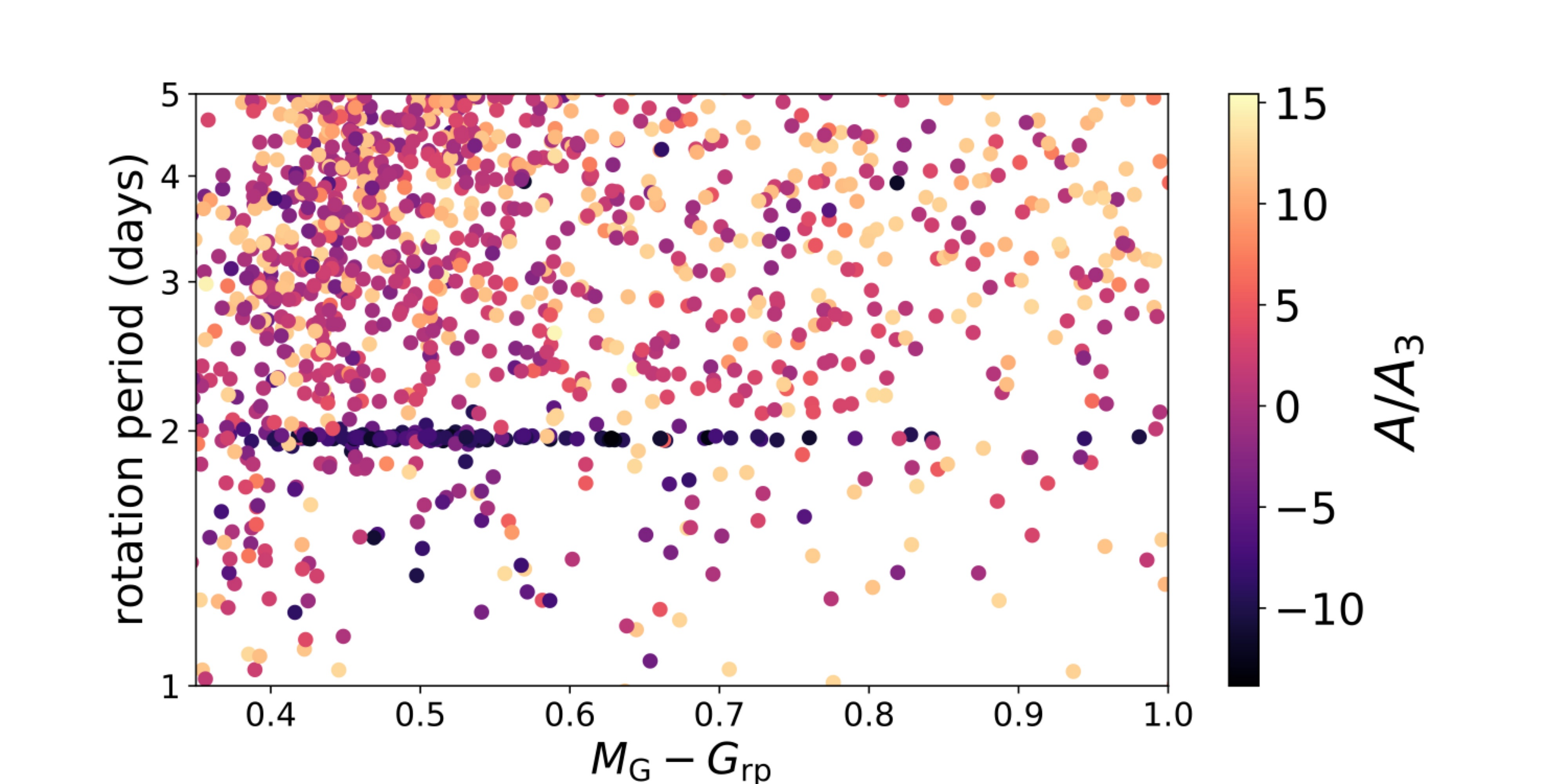}
        \caption{A segment of the period-color diagram showing the presumed artificial pileup at a period of two days. Points are colored by $A/A_3$, the amplitude of the periodic component of the GP relative to the aperiodic component. Stars in the pileup are notable for having a very small value for this ratio relative to the rest of the stars in the sample, allowing us to effectively remove this feature by imposing a cutoff in $A/A_3$ for our final sample.}
        \label{fig:pileup}
    \end{figure}

\begin{figure}
        \includegraphics[width=1.05\hsize]{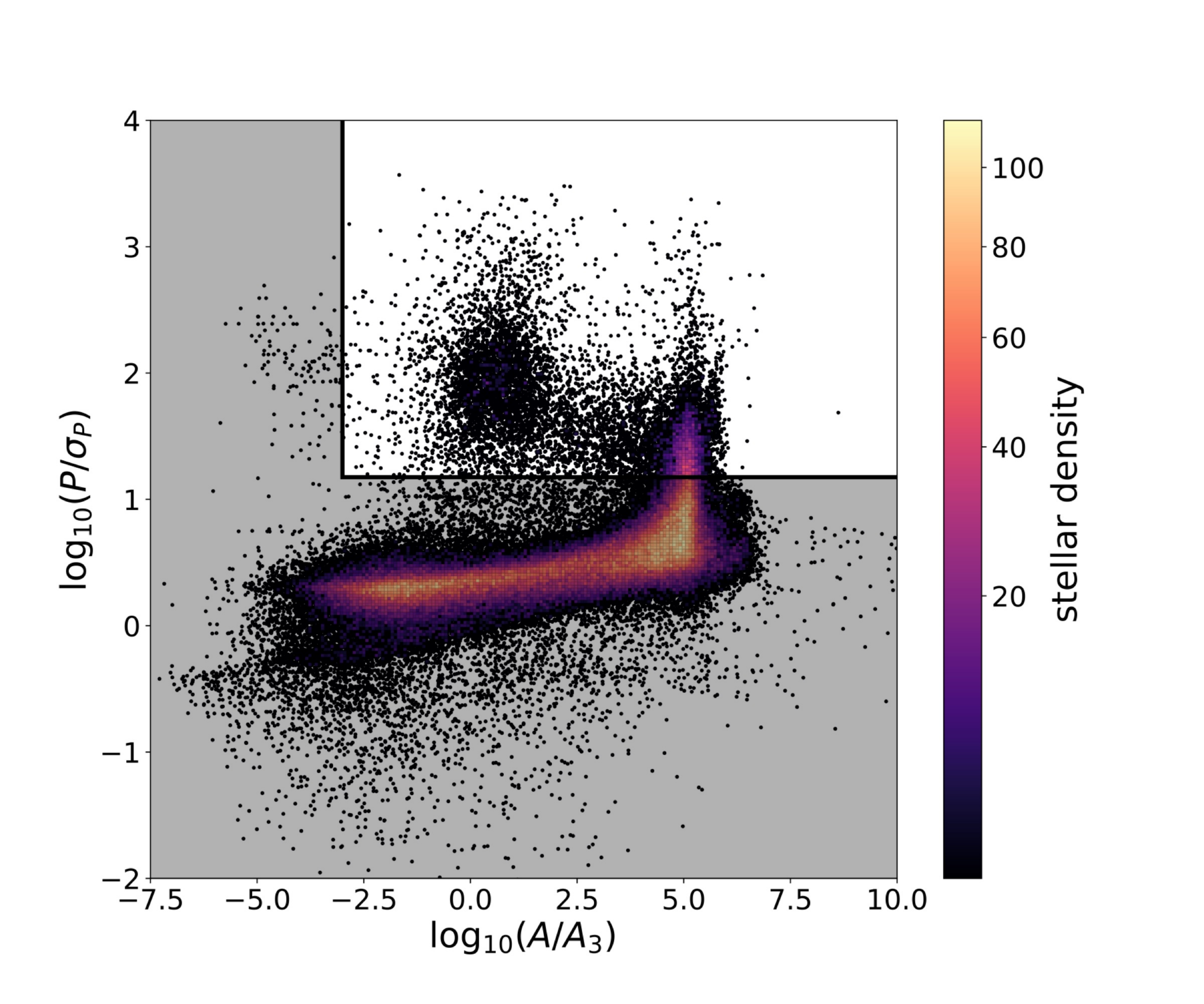}
		\caption{Main-sequence stars plotted in the $P/\sigma_P$ vs $\log(A/A_3)$ plane. The full main-sequence sample is shown with the greyed out area representing the region excluded by the cuts detailed in section \ref{sec:vetting}. Our final sample consists of the stars in the highlighted region.}
		\label{fig:period_amplitude}
	\end{figure}

    We select a final sample of well-measured rotators from the main-sequence sample based on MCMC convergence and period measurement precision. For inclusion of a star in our final sample, we require the following conditions be met:
    
    \begin{itemize}
        \item $P/\sigma_P > 15$
        \item $\log_{10}(A/A_3) > -3$
        \item $0.9 < \hat{R}_P < 1.1$
    \end{itemize}
    where $P$ is the measured period, $\sigma_P$ is the error on the period derived from MCMC, $A$ is the variance of the periodic GP component (the amplitude is $\sigma=\sqrt{A}$, and $A_3 = S_3\omega_3Q_3$ is the amplitude of the aperiodic GP component. $\hat{R}_P$ is the Gelman-Rubin statistic \citep{Gelman1992} which compares the variance of samples for an individual parameter (in this case the period, $P$) within a chain to the variance between chains. For chains that have converged to the same solution, these values will be approximately the same and their ratio, $\hat{R}$, will be close to 1. 
    The cutoff on $\log_{10}(A/A_3)$ is meant to exclude stars for which the periodic component is very small compared to the non-periodic variability on the basis that these stars are more likely to be showing periodicity due to contamination or systematics, rather than rotation. We find that when we do not include this cutoff, a pileup of stars at a period of around two days is observed. This pileup, as shown in Figure \ref{fig:pileup}, spans a range of stellar masses but is extremely localized in period-space and therefore appears to be artificial, though its origin is not known. Figure \ref{fig:period_amplitude} shows our final sample including the main-sequence cuts described in Section \ref{sec:main_sequence} in blue compared to the full main-sequence sample in the $P/\sigma_P$ vs. $\log_{10}(A/A_3)$ plane.
    
    We have chosen these specific conditions with the goal of being conservative about selecting only the highest quality period measurements in order to highlight structure in the color-period diagram. As a result, many periodic signals are excluded from our final sample because they don't meet the condition $P/\sigma_P > 15$. A less selective criteria could be used to obtain a much larger sample size for applications that don't require such high precision, such as an analysis of the rotation period-metallicity relation reported by \citet{Amard2020}. The exclusion of signals on the basis of $P/\sigma_P$ affects the completeness of our sample most at long periods. In Figure \ref{fig:period_color} this means that the upper envelope of the rotation period-color distribution is not as well-defined as it might appear in the plot, especially towards the fainter K and M dwarfs towards the right-hand side of the plot. 
    
\section{Features in period-color space}\label{sec:gap}

\begin{figure*}
        \includegraphics[width=\hsize]{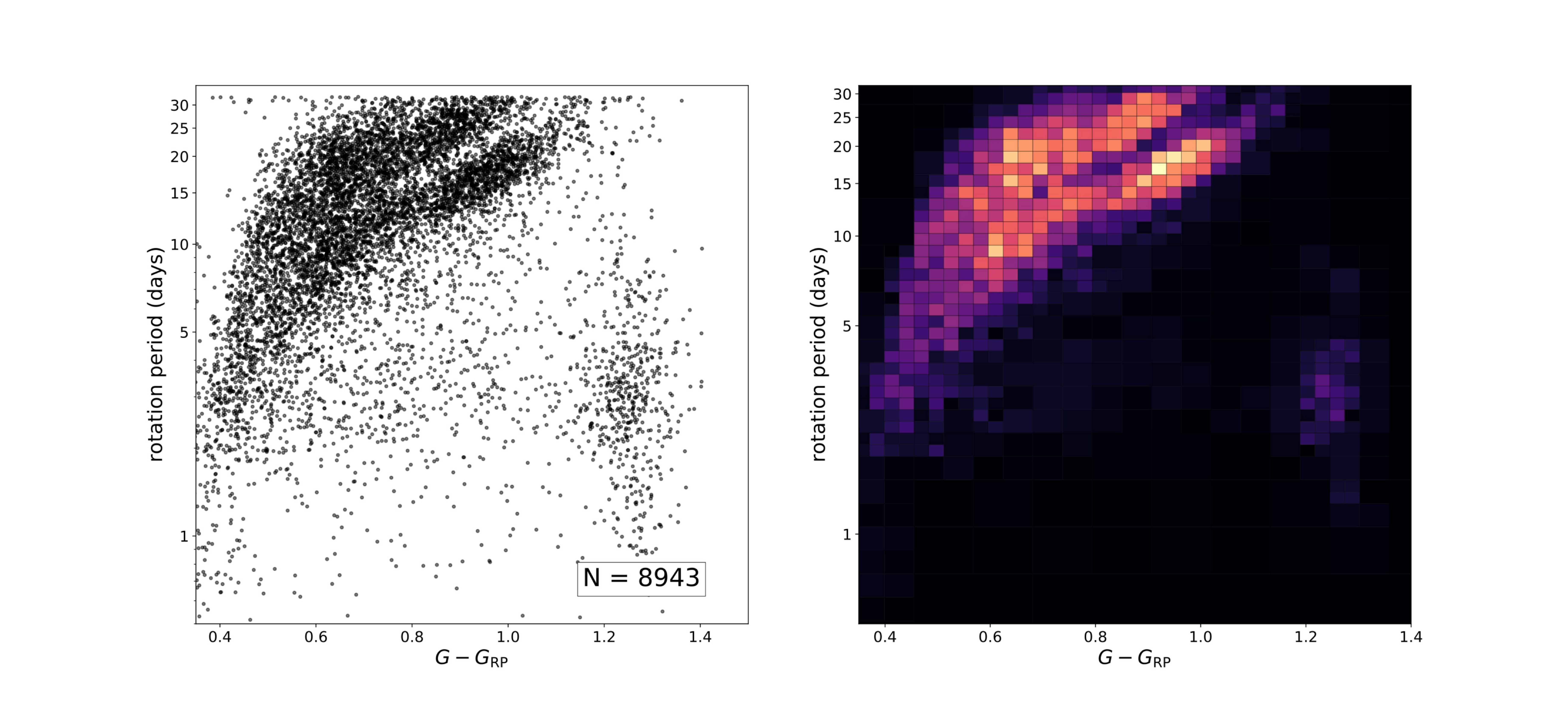}
        \epsscale{1.5}
		\caption{Inferred rotation periods for the 8,943 main sequence K2 stars, plotted against Gaia $G-G_\mathrm{RP}$ color. \textbf{Left:} scatter plot showing the measured periods and \textbf{Right:} The same rotation periods presented as a two-dimensional histogram, in order to highlight the variations in the density of stars across period-color space.}
		\label{fig:period_color}
	\end{figure*}

    Figure \ref{fig:period_color} shows the distribution of rotation periods in period-color space for our final sample. There are several prominent features in this space, among them the afformentioned gap, the upper edge of the envelope of rotation periods, the lower edge of the same envelope, and the overdensity of m dwarfs with short rotation periods in the lower right-hand corner. In this section we focus on the gap and the overdensity of fast-rotating m dwarfs. The upper edge of the envelope is not well-measured in our sample as it is largely determined by the exclusion of rotation periods longer than 32 days, as well as by our cutoff in the precision of the rotation period measurement, which preferentially excludes slow rotators. 
    
    \subsection{The Period Gap}
    The gap in rotation periods extends from $\sim 15$ days at $G-G_\mathrm{RP} = 0.75$ to $\sim25$ days at $G-G_\mathrm{RP} = 1.1$.
	This gap has been extensively studied in the Kepler data, first by \cite{McQuillan2013} for M dwarfs in the Kepler sample. In \cite{McQuillan2014} the same feature was found in Kepler K dwarfs. \cite{Davenport2017} identified the gap for the G dwarfs, and \cite{Davenport2018} showed that the gap is present in stellar populations out to 525 pc, beyond which the feature can no longer be recovered due to the difficulty in recovering rotation periods at large distances. Besides the tentative detection by \cite{Reinhold2020}, our work represents the first robust measurement of this feature outside of the Kepler data. In contrast to the Kepler data, the gap detected here appears wider and has more sharply defined boundaries than is seen in the Kepler sample. 
	
	The robustness of our detection of this gap allows us to constrain new details of the feature. We use an edge detection procedure based on the Canny edge detection algorithm \citep{Canny1986} to find the edges of the period gap. We then fit a parametric model to the gap edges as a function of color. We measure the locations of the gap edge in each campaign individually in order to verify the presence of the gap in each campaign. We confirm that the gap properties do not appear to depend on the direction of the K2 pointing.
	
\begin{figure}
        \includegraphics[width=\hsize]{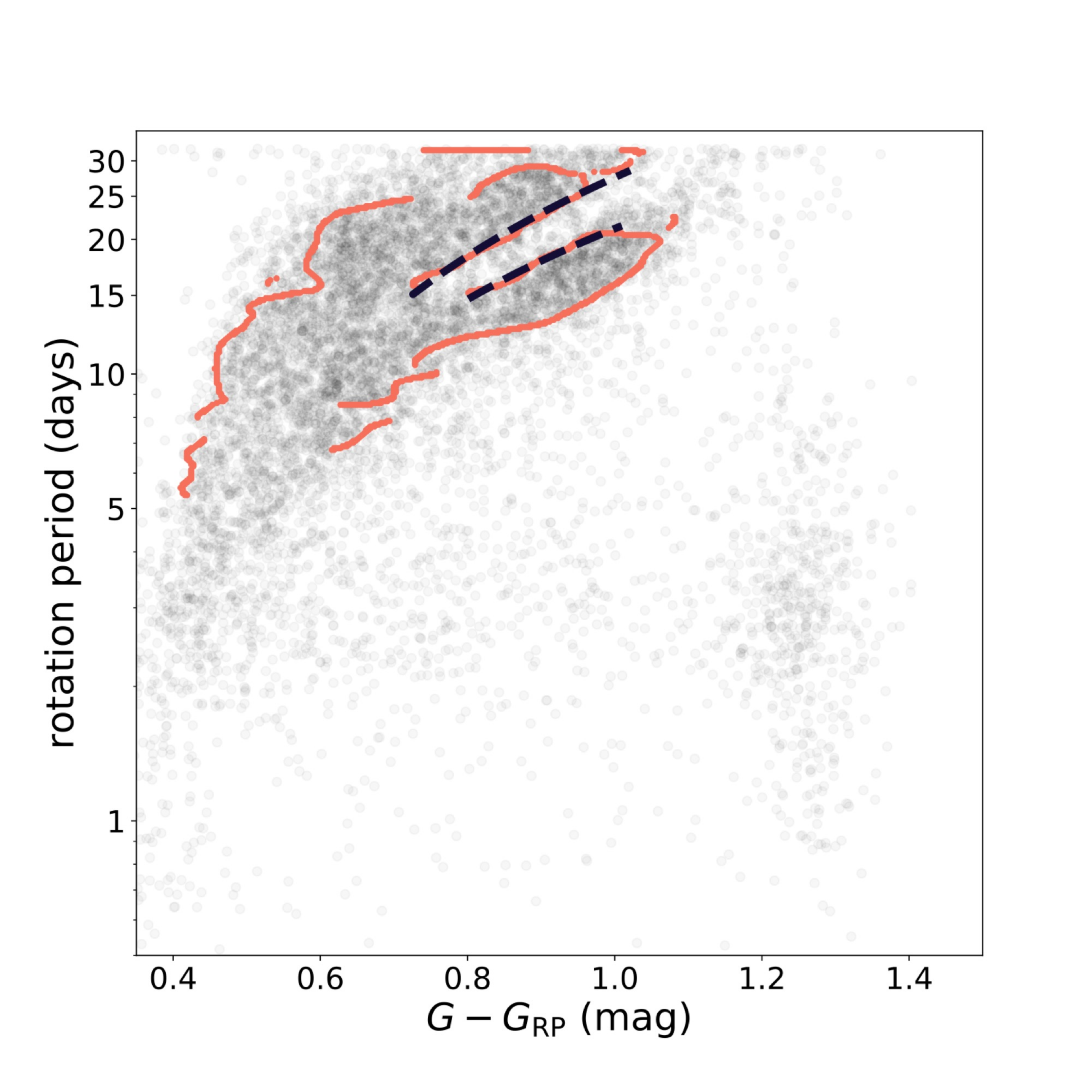}
		\caption{Detected edges of the rotation-period diagram using 
		a modified version of the Canny 
		edge detection algorithm. Best-fit models to the gap edges are shown with blue dashed lines. The model used to fit the edges is given in Equation \ref{eqn:fit} and the best-fit parameters are in Table \ref{tbl:fit}.}
		\label{fig:edge_fits}
	\end{figure}

	We begin by applying our edge detection algorithm to the rotation period-color distribution for all campaigns combined. Our edge detection algorithm is based on the Canny edge detection algorithm, which operates on a two-dimensional array. The Canny algorithm first applies a Sobel operator to the image, which produces an approximation to the gradient. It then identifies local maxima and minima of the approximated gradient which correspond to edges (points where the intensity of the image is changing most quickly) in the original image. Our modification replaces the first step of applying the Sobel operator with the computation of a kernel gradient estimator which allows us to apply the algorithm directly to the distribution of stars in rotation period-color space, which is not possible to do directly with the Canny algorithm as it requires a two-dimensional image grid rather than a set of points in the plane. The kernel gradient estimator is defined to be the gradient of the kernel density estimator. The kernel gradient estimator is defined 
	\begin{equation}
	    \label{eqn:kernel_estimate}
	   \nabla \hat{f}(x, y) = \frac{1}{nh}\sum_i^n\nabla K_i(xh^{-1}, yh^{-1}).
	\end{equation}
	which is the gradient of the more widely known kernel density estimator. $K_i$ is the kernel function taken here to be a Gaussian centered at the coordinates of the $i^\mathrm{th}$ data point, $n$ is the number of data points used to make the estimate, $h$ is the width of the kernel which we set to $0.04$, and $(x, y)$ are the coordinate at which the kernel estimate is computed. We then apply the second part of the Canny algorithm as implemented in \texttt{scikit-image} \citep{scikit-image} to identify local maxima and minima of the kernel gradient estimate. We take these local extrema to be the edges of the distribution. Figure \ref{fig:edge_fits} shows the output of this algorithm applied to our sample. 
	
	We parameterize the gap edges using a function of the form
	\begin{eqnarray}
        \label{eqn:fit}
        P_\mathrm{upper} &=& A(G-G_\mathrm{RP}-x_0)\cr
        &+& B(G-G_\mathrm{RP}-x_0)^{1/2}
    \end{eqnarray}
    using the edges identified in the slice of color-space given by 
	\begin{equation}
	    0.8 < G - G_\mathrm{RP} < 1.05.
	\end{equation}
	which corresponds to the stellar mass range:
	\begin{equation}
	    0.57M_\odot < M < 0.76M_\odot.
	\end{equation}
	Equation \ref{eqn:fit} is taken from the gyrochronology model of \cite{Barnes2003}. Our decision to fit the gap edges with this equation is motivated by the observation that the gap edges appear to have a similar trend to the gyrochrones from that work, but this choice is not meant to imply that the gap edges occur at constant age. The best-fit parameters are given in Table \ref{tbl:fit}, and Figure \ref{fig:edge_fits} shows the best-fit models in period-color space. 
	
	\begin{deluxetable}{l|cccc}
	\label{tbl:fit}
	\tablenum{7}
	\tablecaption{Best fit parameters from Equation 
	\ref{eqn:fit}.}
	\tablehead{\colhead{} & \colhead{A (days)} & \colhead{B (days)} & \colhead{$x_0$}} 
        \startdata
        upper edge & 68.2277 & -43.7301 & -0.0653 \\
        lower edge & 34.0405 &  -2.6183 & 0.3150
        \enddata
    \end{deluxetable}
    
    \begin{deluxetable*}{c|ccc}
	\tablecaption{Measured gap edges and widths.}
	\label{tbl:gap}
	\tablehead{\colhead{$G_\mathrm{BP}-G_\mathrm{RP}$ (mag)} & \colhead{$P_\mathrm{lower}$ (days)} & \colhead{$P_\mathrm{upper}$ (days)} & \colhead{ gap width (days)}} 
        \startdata
            0.80 & 15.20 & 17.97 & 2.771 \\
            0.85 & 15.92 & 19.90 & 3.98 \\
            0.90 & 17.97 & 22.26 & 4.29 \\
            0.95 & 19.54 & 24.89 & 5.35 \\
            1.00 & 20.66 & 28.61 & 7.95 
        \enddata
    \end{deluxetable*}
	
\begin{figure*}
        \plottwo{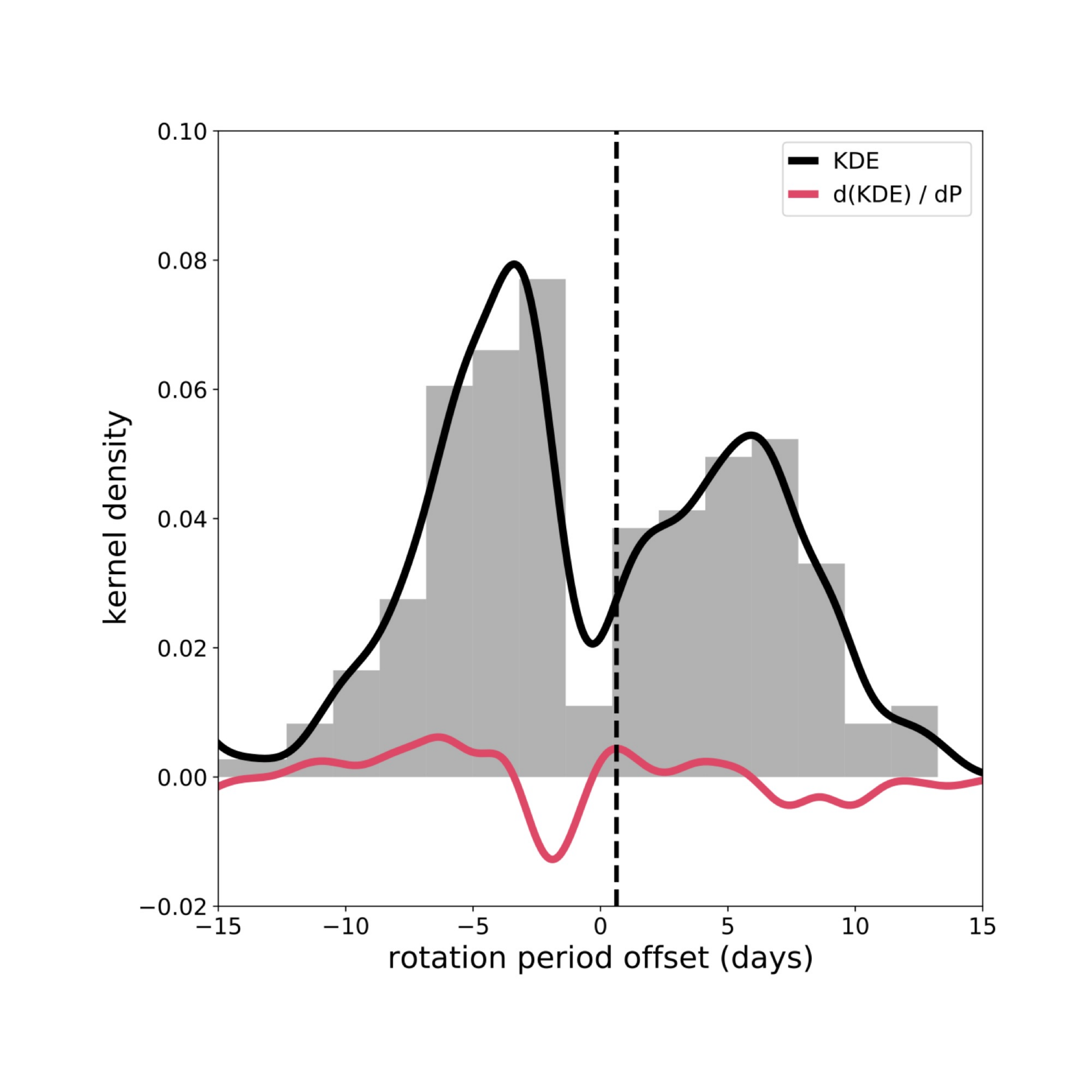}{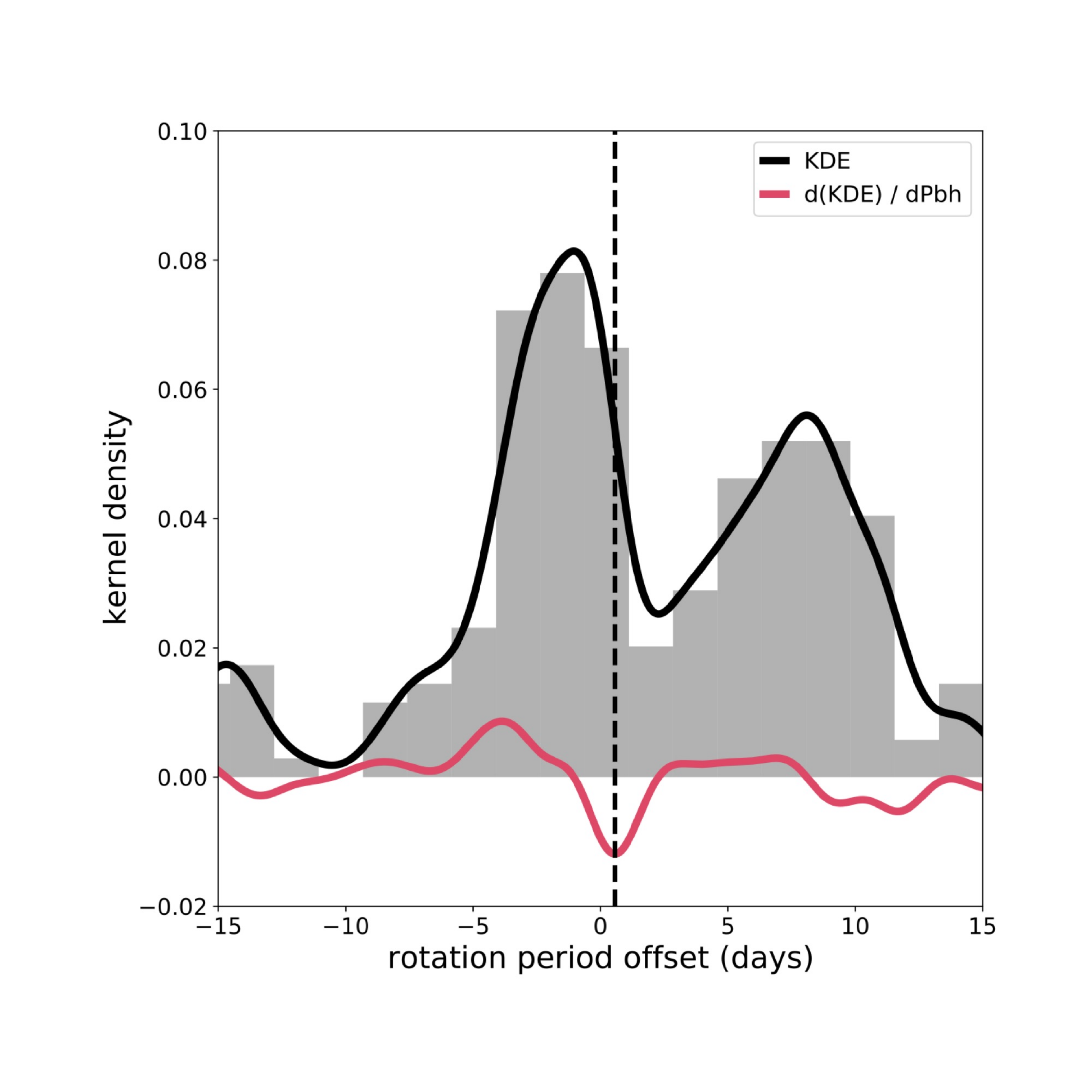}
		\caption{Histograms showing the distribution of rotation periods for stars in Campaign 8, in the color grange $0.8 < G-G_\mathrm{RP} < 1.05$. The kernel density estimate is shown in black, and its derivative is shown in red. The locations determined for the gap edges are shown by the dashed vertical line. The left panel shows the upper gap edge and the periods are given as the difference between the observed rotation period and the trend of the upper gap edge. The right panel shows the same for the lower gap edge.}
		\label{fig:peaks}
	\end{figure*}

	We have estimated the locations of the gap edges in each Campaign individually in terms of their offset from the best-fit model for the full sample. Because the sample sizes are small for some campaigns we collapse the problem to one dimensions rather than considering the full two-dimensional period-color diagram. For each edge of the gap (upper and lower) in each Campaign, we first subtract off the gap trend and then sum over color in the range of colors for which the best-fit edge model is valid:
	\begin{equation}
	    0.8 < G - G_\mathrm{RP} < 1.05.
	    \label{eqn:color_slice}
	\end{equation}
	This gives us the one-dimensional period distribution in terms of the offset from the edge locations defined in table \ref{tbl:fit}. Treating each gap edge separately, we apply the Gaussian kernel derivative estimator, which is the one-dimensional anologue of Equation \ref{eqn:kernel_estimate}, to the period distribution. We then identify the local maximum (in the case of the upper edge) or minimum (in the case of the lower edge) nearest to the zero-offset point and take this to be an estimate of the location of the edge relative the edges in Figure \ref{fig:edge_fits}. Figure \ref{fig:peaks} illustrates this procedure for Campaign 8. 
	
	Figure \ref{fig:edges_campaign} shows the lower edge locations plotted against the upper edge locations for 16 of the 18 campaigns. We have excluded Campaigns 0, 2, and 11 which have too few stars to make an accurate determination of the edge locations. The error bars are determined by bootstrap resampling from the full sample. We do not observe an obvious correlation between the upper and lower gap edges, which would indicate a shifting of the gap towards longer or shorter periods for some campaigns. There are a few outlier campaigns, with Campaign 18 being the most significant. Campaigns 4, 8, 10, 13, and 15 also deviate noticeably from at least one of the measured gap edges for the all-campaign sample. In Figure \ref{fig:period_color_campaign} we show the period-color diagrams for these outliers. We note that Campaigns 5 and 18 observed the Praesepe cluster which imprints a visible sequence of stars corresponding to a 600-700 Myr gyrochrone onto the diagram. This does not appear to influence the gap measurement for Campaign 5, but for Campaign 18 the Praesepe cluster is likely responsible for the displacement of the lower gap edge from the expected position. For the rest of the outliers we don't see any obvious evidence of a systematic displacement in the gap edges from the locations derived from the all-campaign data. In general we don't find the outliers to be significant, and we conclude that the rotation period gap shows no dependence on the direction of the K2 pointing. 
	
	To further illustrate this point, Figure \ref{fig:map} shows the locations of the K2 footprints for these outlying campaigns on the sky relative to the rest of the campaigns. We note that the outlier campaigns are in general widely separated on the sky, appear both above and below the galactic plane, and show no evidence of clumping. We therefore state our conclusion that the rotation period gap appears to be an isotropic feature of the stellar populations in the nearby Milky Way.
	
	\begin{figure}
	    \centering
	    \includegraphics[width=\hsize]{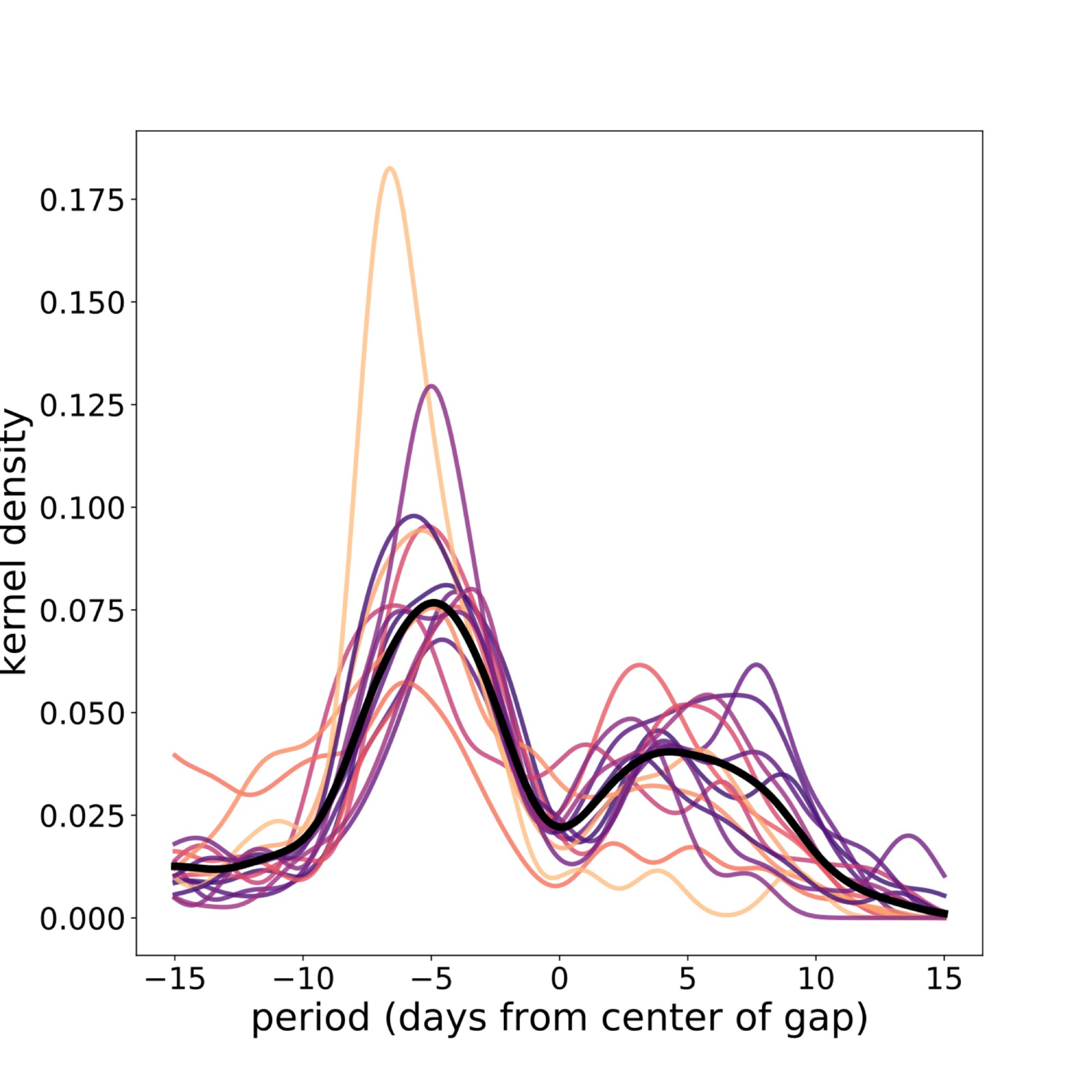}
	    \caption{Kernel density estimates for the 16 campaigns with $N > 200$, for stars with $G-G_\mathrm{RP}$ in the range defined in equation \ref{eqn:color_slice}. The thick black curve is the kernel density estimate for all 16 campaigns combined.}
	    \label{fig:kde_campaigns}
	\end{figure}
	
	\begin{figure}
	    \centering
	    \includegraphics[width=\hsize]{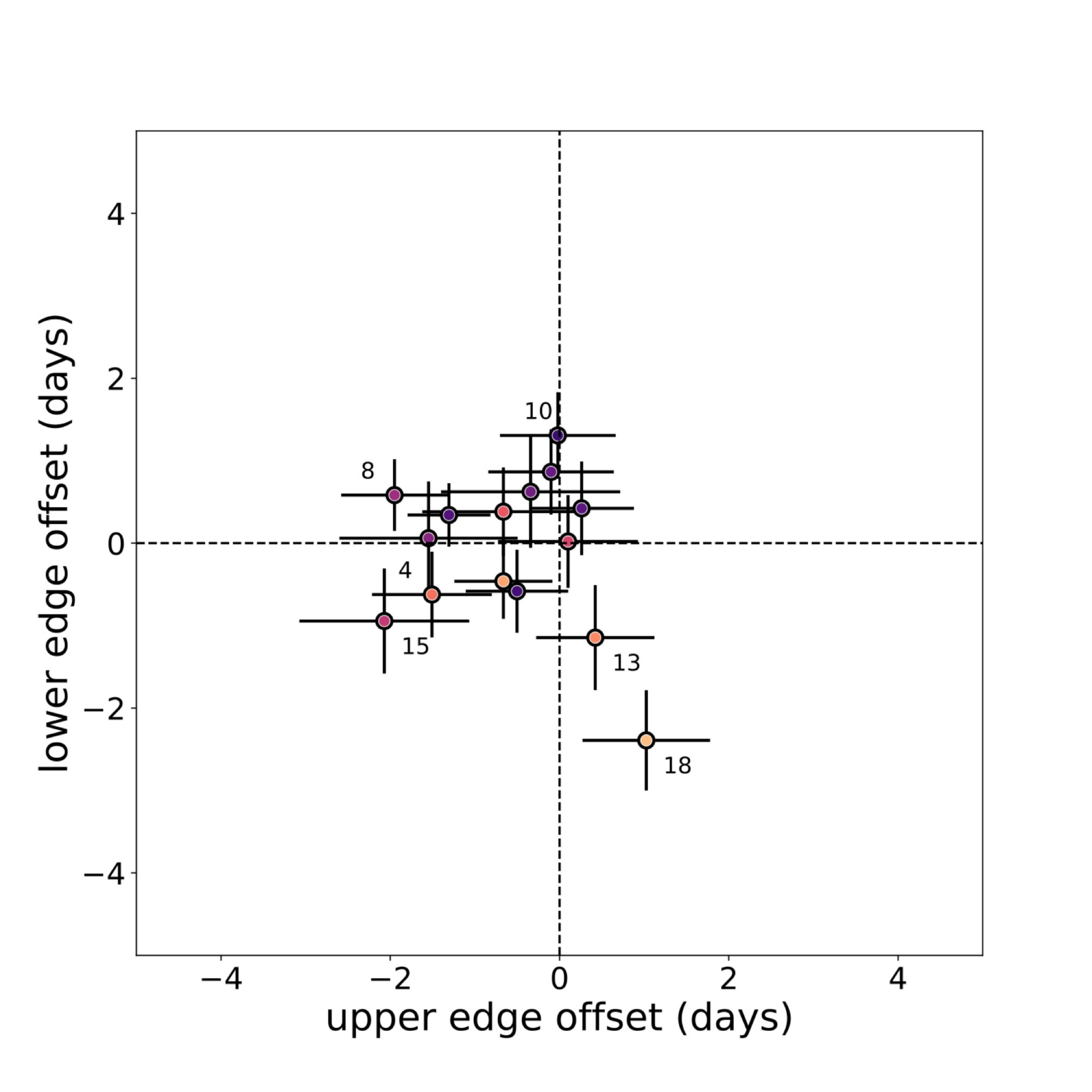}
	    \caption{Location of the lower edge of the gap plotted 
	    against the location of the upper edge. Both gap edge 
	    locations are given as the displacement in days from the 
	    best-fit gap edges in Figure \ref{fig:edge_fits} and 
	    are determined from the slope of the one-dimensional kernel 
	    density estimates in figure \ref{fig:kde_campaigns}. The four most 
	    notable outliers are labeled and their period-color 
	    diagrams are shown in figure \ref{fig:period_color_campaign}.}
	    \label{fig:edges_campaign}
	\end{figure}
	
	\begin{figure*}
	    \plottwo{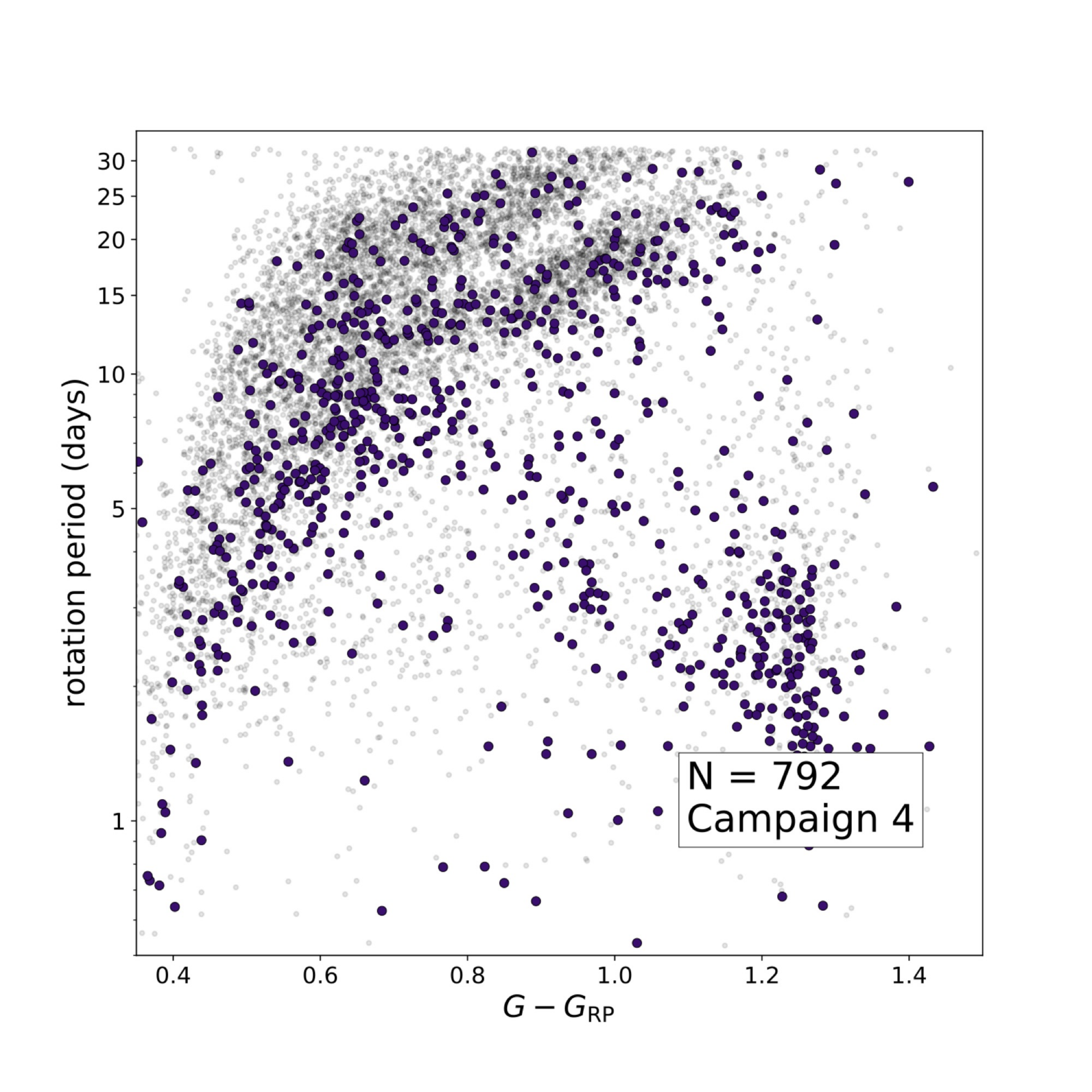}{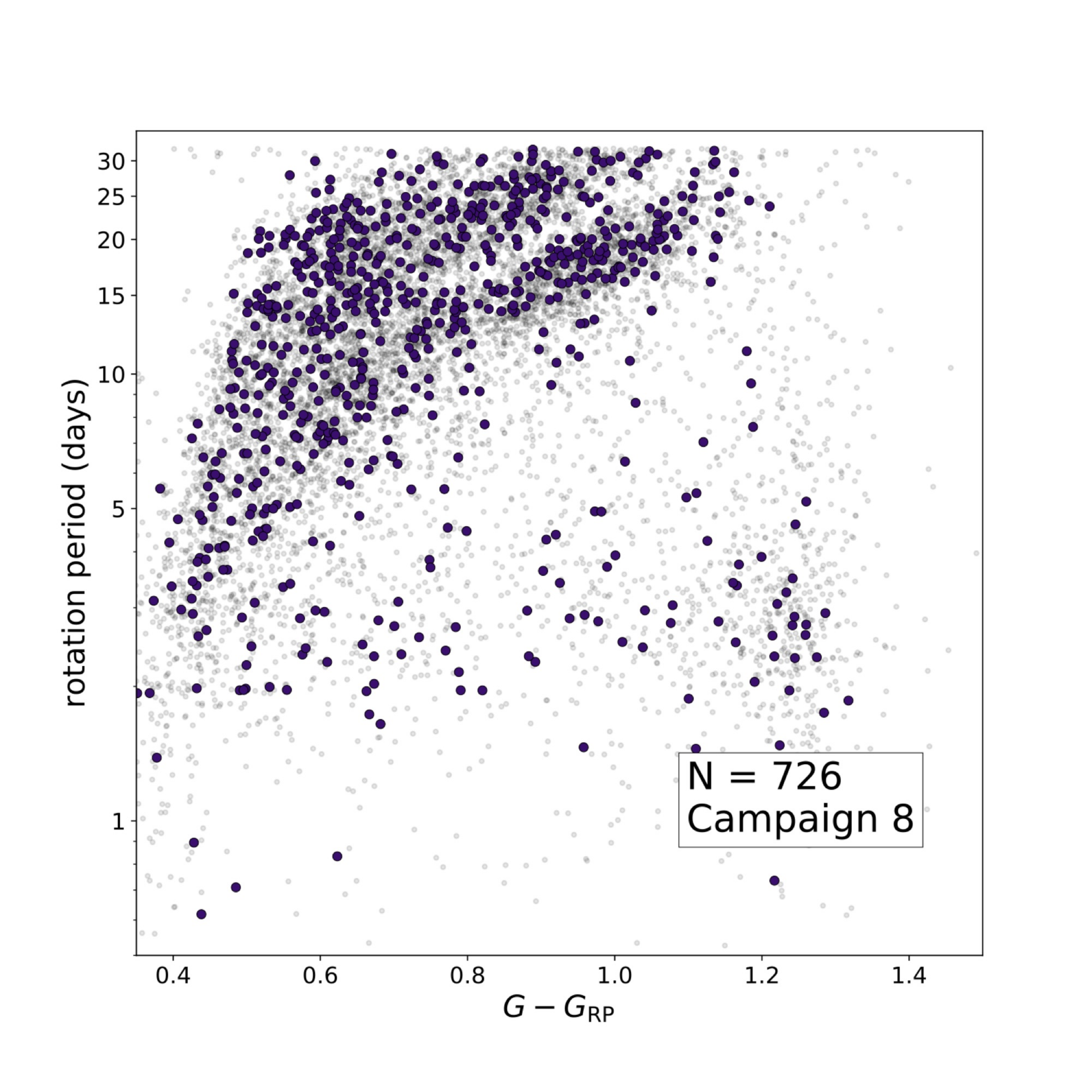}
	    \plottwo{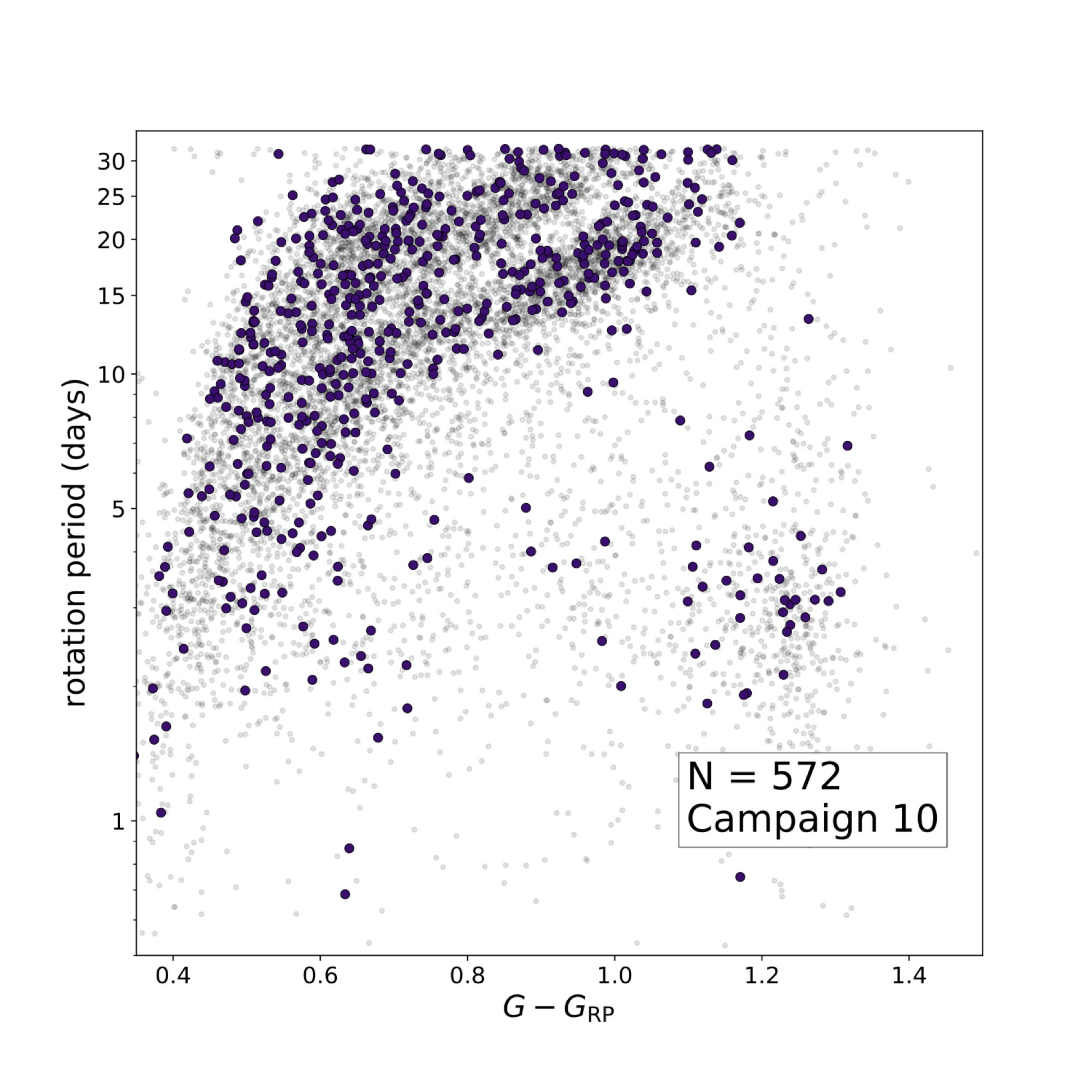}{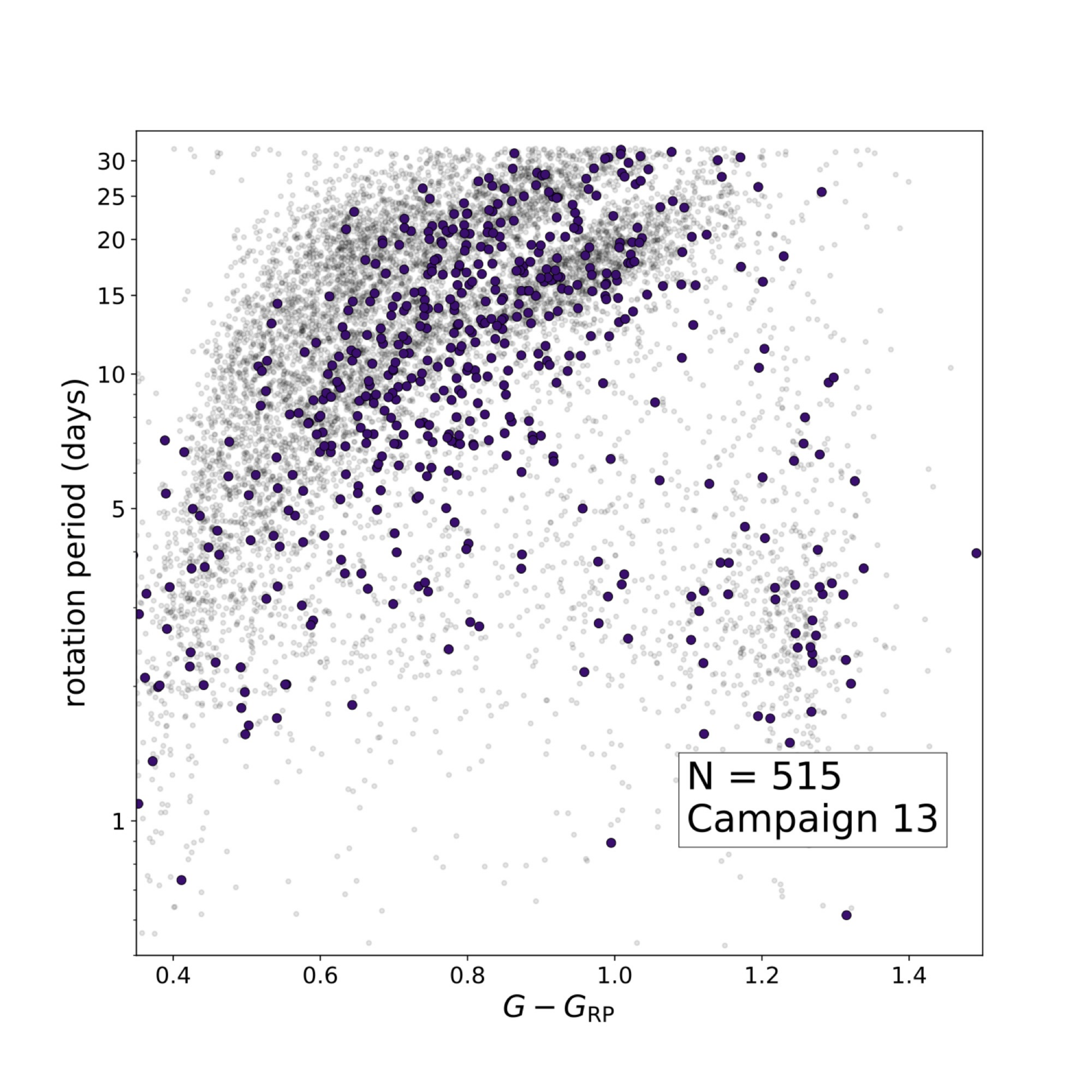}
	    \plottwo{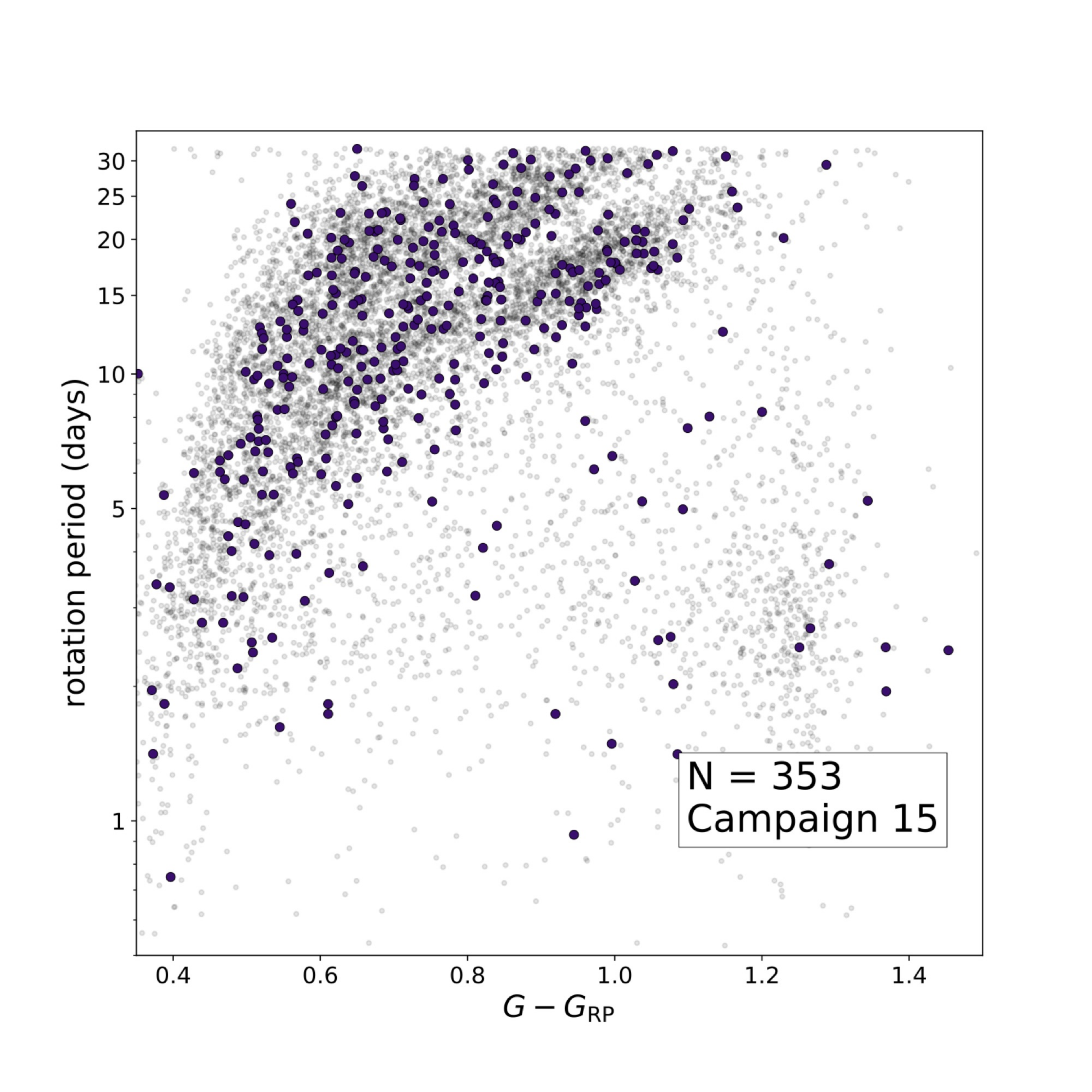}{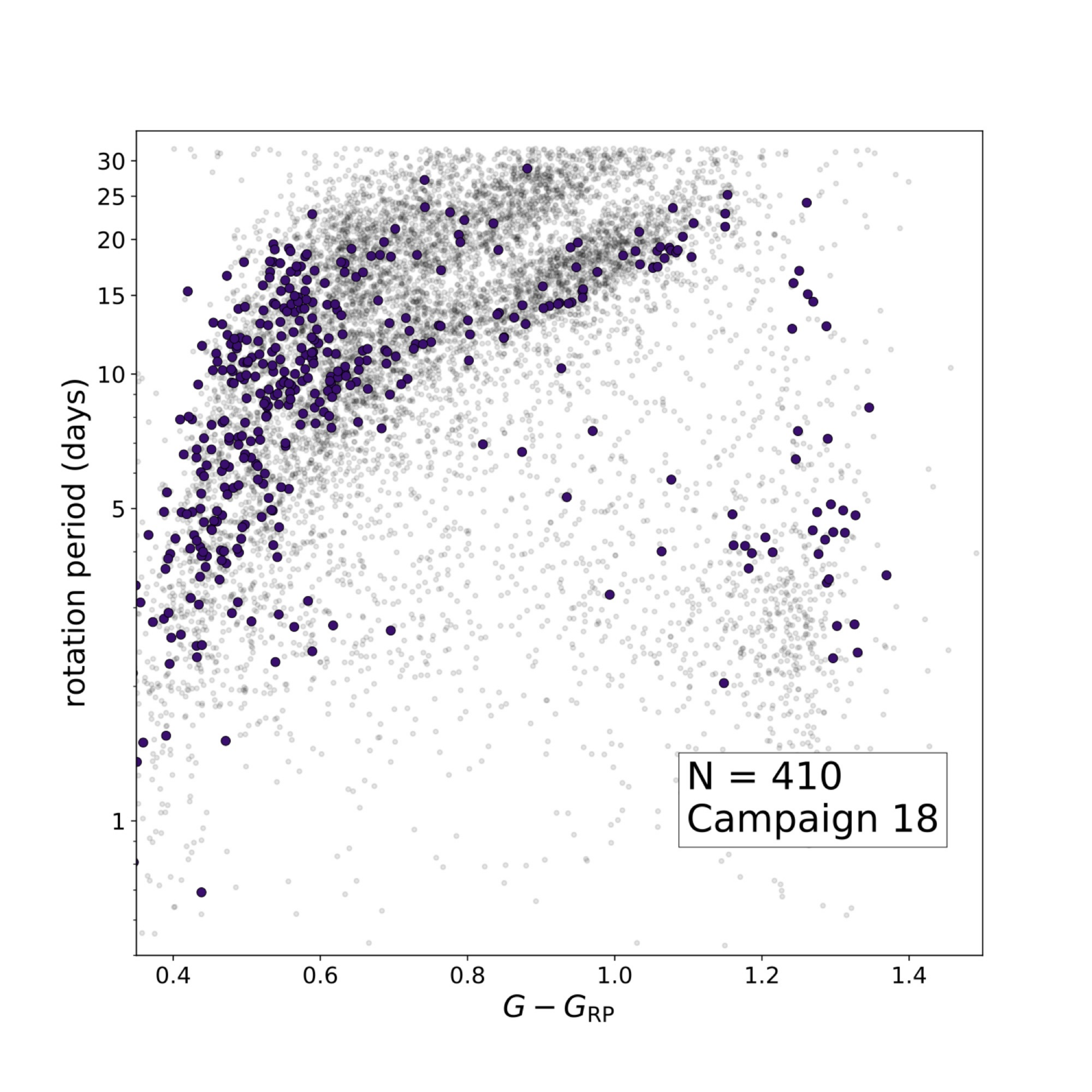}
	    \caption{Period-color plots for the 
	    outliers identified in Figure \ref{fig:edges_campaign}. The dark purple points show the 
	    stars from the individual campaign, while 
	    the gray points are the full sample 
	    of 8,943 stars. The locations of the K2 footprints on the sky for these campaigns is shown in Figure \ref{fig:map}. The Praesepe sequence can be seen in Campaign 18, but it only appears to impact the the gap edge detection for Campaign 18, as seen in Figure \ref{fig:edges_campaign}. With the possible exception of Campaign 13, the gap appears to be respected by the subsamples for each campaign, which indicates to us that the outliers in Figure \ref{fig:edges_campaign} are the result of stochastic variations within the sample and are not significant.}
	    \label{fig:period_color_campaign}
	\end{figure*}
	
	\begin{figure*}
        \plotone{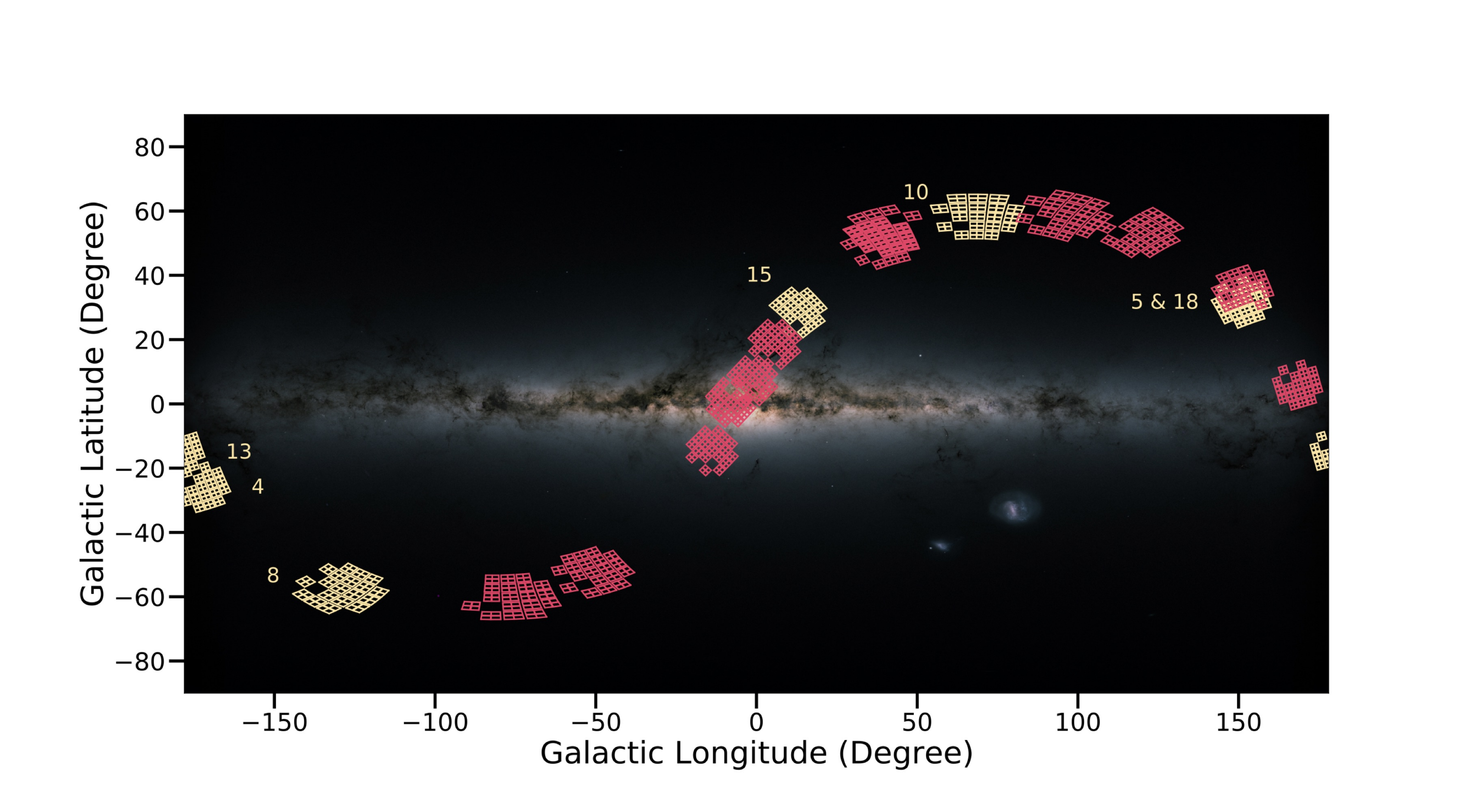}
    		\caption{Positions of the campaigns shown 
		in Figure \ref{fig:period_color_campaign} 
		with Milky Way as seen by Gaia DR2 for reference. There is no obvious correlation between the direction of the K2 pointing and the change in shape or position of the gap for the outlier campaigns. Background image credit: ESA/Gaia/DPAC.}
		\label{fig:map}
	\end{figure*}
	
	\subsection{Other features in the rotation period-color diagram}
	
	In addition to the prominent period gap, a major feature of the period-color distribution seen here is the over-density in the bottom left corner, which represents a population of fast rotating M-dwarfs. This population has been studied in young open clusters \citep{Rebull2016, Rebull2018} as well as in the MEarth sample \citep{Newton2016}. The break between the slow rotating M dwarfs for which the rotation period increases with decreasing mass, and the fast-rotating sequence for which the rotation period decreases with decreasing mass occurs at approximately the mass at which M dwarfs become fully convective and corresponds to a change in the morphology of the surface magnetic field from a more complex towards a more simple configuration \citep{Morin2010, Garraffo2018}.
	
\begin{figure}
        \includegraphics[width=\hsize]{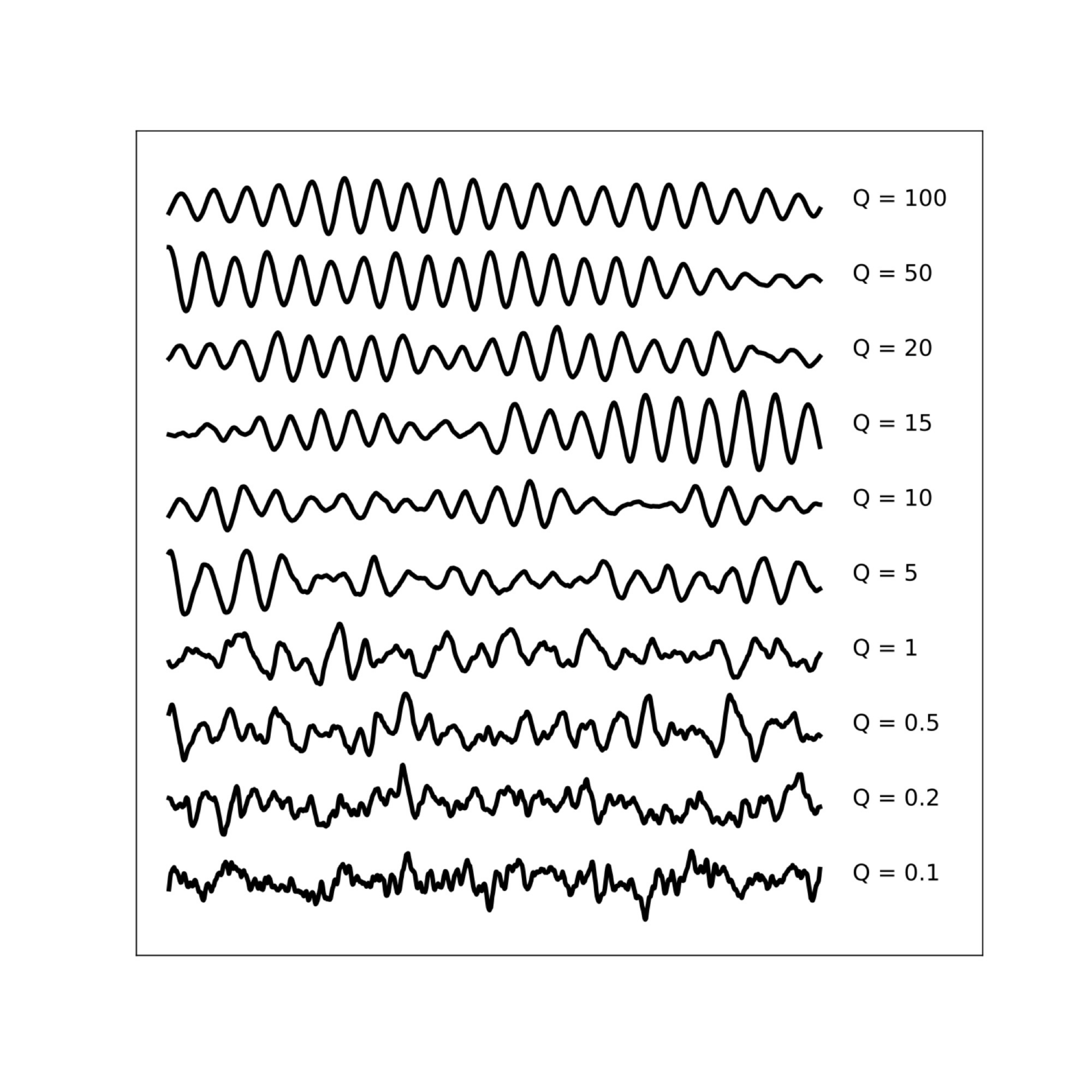}
    		\caption{Samples drawn from our GP model showing the effect of increasing the quality factor, $Q$, on a light curve. All light curves have the same period and amplitude. We have set $\Delta Q = 0$ for these simulations so that $Q = Q_\mathrm{max}$. A higher $Q$ value means that the light curve shows stronger periodicity. In terms of stellar rotation, this likely indicates that surface features are stable for a longer period of time when $Q$ is large.}
		\label{fig:light_curves_Q}
	\end{figure}

	We find that the light curves of these fast-rotating M dwarfs show rotational modulation that is more periodic than those of the other stars in our sample. For our GP rotation model the degree to which a light curves shows periodic variations is measured by the parameters $Q_1$ and $Q_2$ in equation \ref{eqn:kernel}. Larger values of $Q_1$ and $Q_2$ mean that the power spectrum of the variability is more sharply peaked about $\omega_1$ and $\omega_2$. In this analysis we consider the maximum of $(Q_1, Q_2)$ which we call $Q_\mathrm{max}$. Figure \ref{fig:light_curves_Q} demonstrates how this parameter effects the appearance of the light curve. We interpret a larger $Q_\mathrm{max}$ value to indicate that features on the star's surface are stable over a longer period of time, giving rise to variations that are coherent across many periods. 
	
\begin{figure*}
        \includegraphics[width=\hsize]{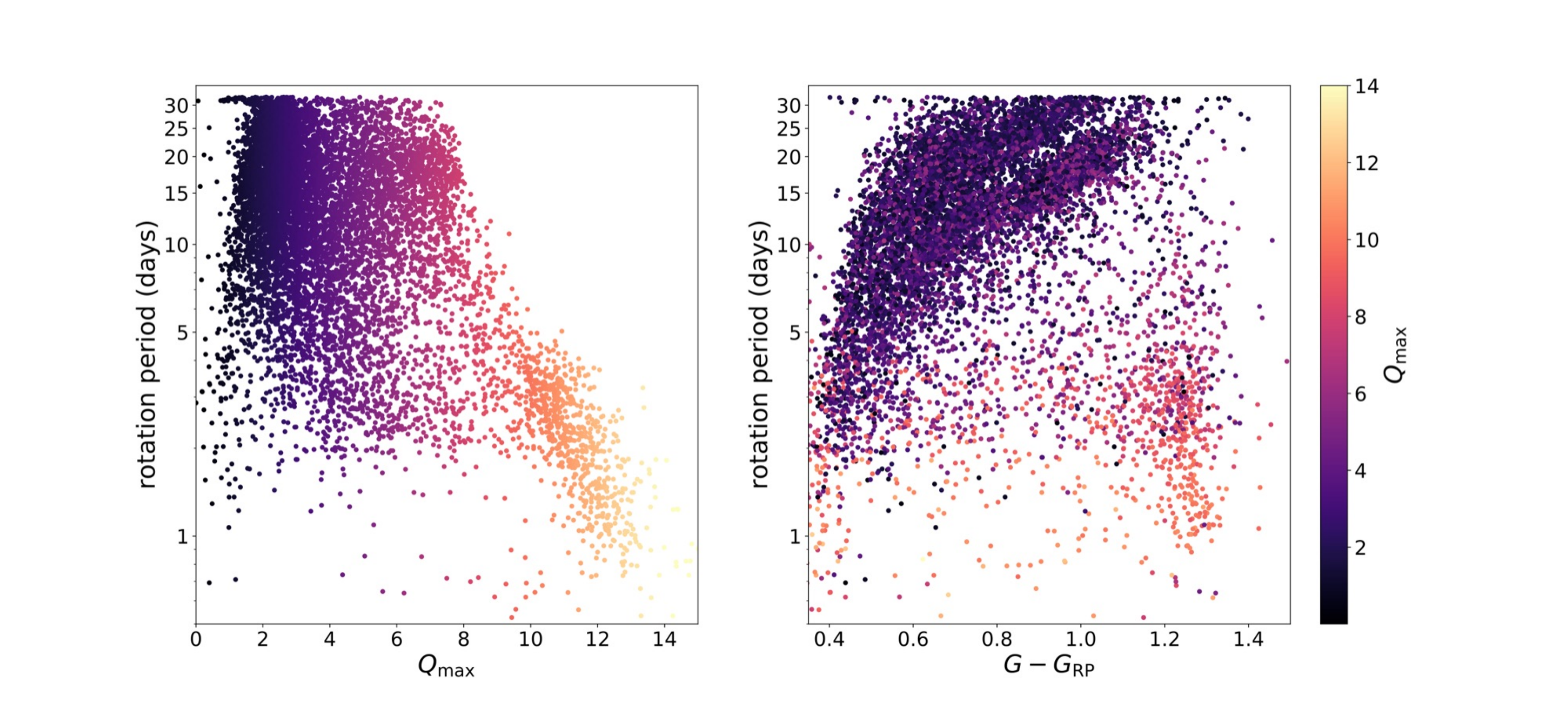}
    		\caption{\textbf{Left:} Sample in period-$Q_\mathrm{max}$ space, showing that sinusoidal rotators (high-Q stars) cluster separately from the main population and preferentially occur at short rotation periods. \textbf{right:} Sample in period-color space with stars colored by the maximum quality factor $Q_\mathrm{max}$. While stars with higher $Q_\mathrm{max}$ values cluster in the fast-rotating M dwarfs, they also occur across all colors, and hence across all stellar masses in our sample.}
		\label{fig:period_color_Q}
	\end{figure*}

	The right panel of Figure \ref{fig:period_color_Q} shows how $Q_\mathrm{max}$ varies across the period-color diagram. From this figure we see that stars with large $Q_\mathrm{max}$, indicating stronger periodicity, cluster in the fast-rotating M dwarfs. These stars are, however, not limited to this cluster and occur in lower densities across the full range of $G-G_\mathrm{RP}$ at short rotation periods. The left panel of Figure \ref{fig:period_color_Q} shows our sample in period-$Q_\mathrm{max}$ space with stars colored by their $G-G_\mathrm{RP}$ color. In this space we observe a distinct population of strongly periodic rotators with a negative correlation between rotation period and $Q_\mathrm{max}$. 
	
\section{Discussion}\label{sec:discussion}

    There has been much interest in, and discussion of, the origin of the rotation period gap, with several promising possible explanations having been put forward since its discovery \citep{Angus2020, Davenport2018, Reinhold2013, Reinhold2019}. We now consider these potential explanations in light of our new measurements, as well as taking into account recent work on the 2.7 Gyr cluster Ruprecht 147 by \cite{Curtis2020} and \cite{Gruner2020}, which crosses the rotation period gap. 
    
    \cite{McQuillan2014} and \cite{Davenport2018} propose that the gap may be an artifact of a recent ($< 500 $ Myr) burst of star formation in either the Solar neighborhood or in the direction of the Kepler field, which would have produced a population of young, fast-rotating stars that make up lower branch of the observed bimodality. The single pointing of the Kepler mission admitted the possibility that this feature confined to that field. Our sample has the benefit of K2's multiple pointings, which has allowed us to demonstrate that the bimodality is present in all directions and is therefore not unique to the stellar population observed by Kepler. The possibility remains that the bimodal star formation history suggested by \cite{McQuillan2014} and \cite{Davenport2018} might be isotropic. However, the position and shape of the gap revealed by our sample makes this explanation untenable, as the trend of the gap shows a sharper slope than the sequences associated with constant age populations from Praesepe and NGC 6811 \citep[e.g.][]{Curtis2019}.  It is interesting to note that the sequence of stars associated with the 2.7 Gyr cluster Ruprecht 147 appears to cross the gap around $G - G_\mathrm{RP} \sim 0.7$. While caution should be exercised due to the fact that this sequence has a relatively smaller number of stars than for the younger clusters and shows a large intrinsic scatter, this apparent crossing of the gap lends further evidence against the hypothesis that the gap represents a feature at constant age. Indeed, as \citet{Curtis2020} note in their analysis of periods for Ruprecht 147 and other clusters, this gap-crossing seems to occur at a roughly fixed Rossby number, rather than at a single fixed age, which agrees with our assessment.
    
	\cite{Reinhold2013} and \cite{Reinhold2019} suggest that the gap is an artifact of the transition from spot-dominated to faculae-dominated photospheres as stars age. In this explanation the gap would result from a minimum in the detectability of rotation periods for stars at the point in this transition where neither spot nor facula-induced variability are able to dominate the light curves of these stars. Our measurements of the gap do not rule out this explanation. To do this would require more work on the evolution of stellar activity over a range of ages and spectral types. 
	
	Our preferred explanation is that the gap emerges from a period of accelerated spindown immediately after the stalled spindown noted by \cite{Curtis2020}. This explanation was first put forth by \cite{McQuillan2013}, but the hypothesis was dismissed in favor of the ``two populations'' hypothesis preferred by \cite{McQuillan2014} and \cite{Davenport2018}.
	
\begin{figure*}
        \includegraphics[width=\hsize]{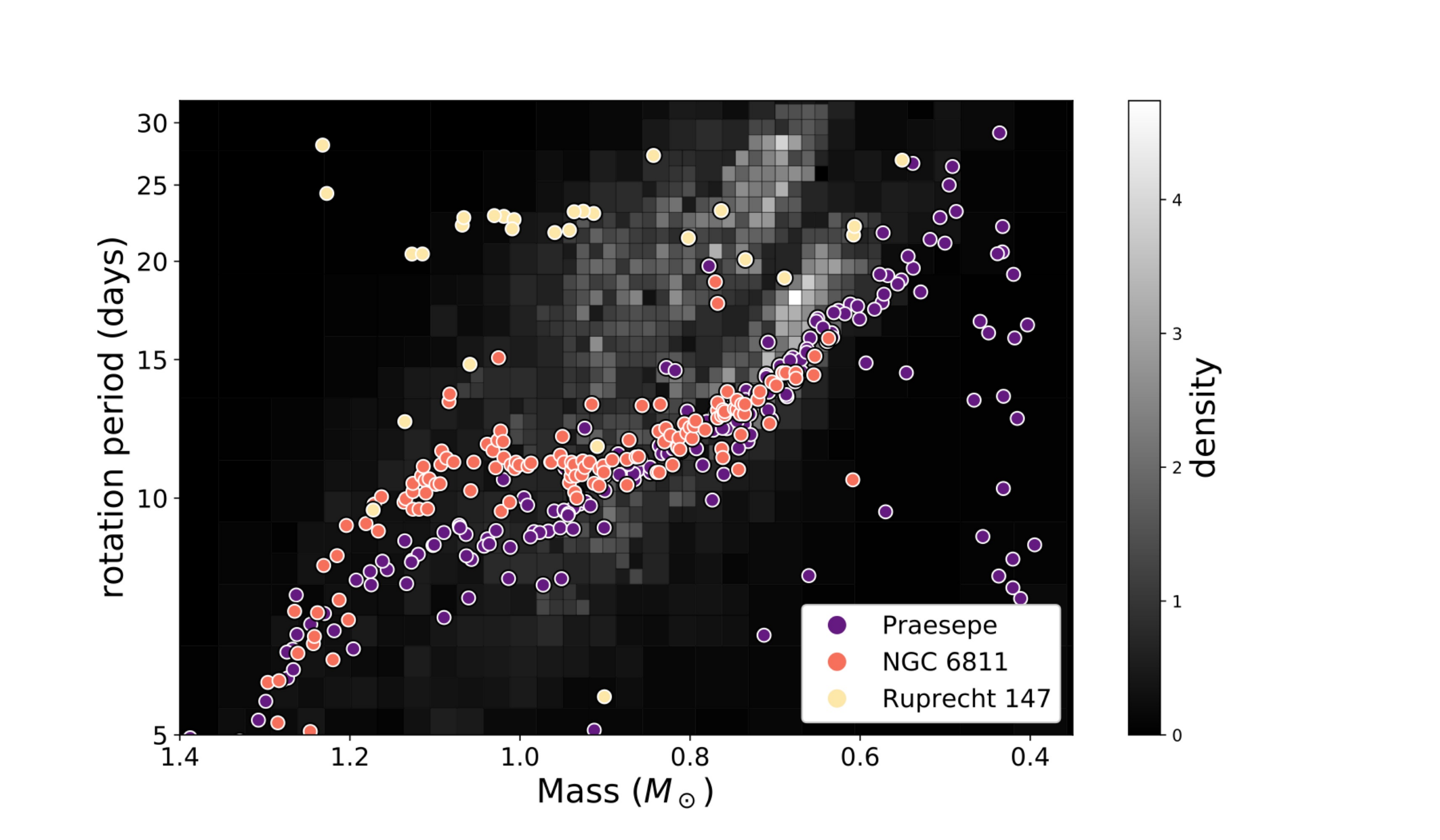}
    		\caption{Clusters Praesepe, NGC 6811, and Ruprecht 147 superimposed on the distribution of field stars. For Praesepe, we use our own rotation measurements with membership in the cluster taken from \cite{Douglas2019}. Rotation periods for NGC 6811 are from \cite{Curtis2019}, and for Ruprecht 147 from \cite{Curtis2020}.}
		\label{fig:clusters}
	\end{figure*}

	In this scenario, a young star with its envelope initially decoupled from its core would experience magnetic braking, reducing the spin of the envelope while the decoupled core would be allowed to continue its faster rotation. At a later time the core and envelope would begin to exchange angular momentum. At this point the transfer of angular momentum from the core to the envelope would slow or even halt spindown by offsetting magnetic braking at the surface, resulting in an overdensity of stars just below the period gap. The underdensity making up the gap itself could then be explained by a period of increased spindown once this coupling is complete and before the star resumes ordinary Skumanich spindown. This could be due to a temporary increase in magnetic activity. \cite{Lanzafame2015} and \cite{Spada2020} have developed a spindown model featuring a mass-dependent core-envelope coupling timescale which reproduces the stalling behavior and has been applied to observations of open clusters. \citet{Curtis2020} have found the stall in spindown corresponds to a track of roughly constant Rossby number.  \cite{Angus2020} suggests that this mechanism may explain the period gap as a break between a ``young'' regime in which rotation periods increase with decreasing mass from an ``old'' regime in which rotation periods are nearly constant or even decreasing with decreasing mass, with the gap representing a period of relatively fast spin evolution during the transition between these regimes. In Figure \ref{fig:clusters} we plot rotation measurements for several important clusters over the distribution of K2 field stars to show the correspondence between the location of the gap in the field stars and the apparent stalling of spindown in clusters. We use the open clusters Praesepe, which has an age of 600-700 Myrs, NGC 6811 which has an age of approximately 1 Gyr \citep{Agueros2018, Curtis2019}, and Ruprecht 147, which is older than both at 2.7 Gyrs. The sequences for Praesepe and NGC 6811 sit on top of each other for low mass stars, but have diverged for stars more massive than about 0.9 $M_\odot$, suggesting that lower mass stars have stalled in their angular momentum loss while higher mass stars have continued to spin down. By the time we reach the age of Ruprecht 147, spindown has resumed for stars down to about $0.7 M_\odot$. The location of the gap in the rotation measurements for the K2 field stars coincides with the point at which the clusters transition from stalled spindown at low masses to resumed Skumanich spindown at higher masses. 
	This supports the notion that the gap represents a discontinuity between these two regimes of spindown.
	
	There is still much work to be done to determine whether core-envelope coupling and decoupling fully explains these observations. On the observational side it will be important to continue to benchmark clusters of ages between 1 Gyr and solar age, as clusters in this age range may cross the gap (similar to Ruprecht 147). On the theoretical side, models of rotational evolution that can explain the period of rapid spindown after the epoch of stalled spindown and in doing so reproduce the shape and trend of the gap will be important for testing this explanation. Another promising avenue of investigation may be kinematic dating of field stars \citep{Angus2020}. Proper motion measurements from Gaia may provide us with the ability to estimate ages of stellar populations by the vertical component of their motion with respect to the galactic disc, since stars become excited in this direction by dynamical interactions over time. This would allow for independent calibration of gyrochronological relations, which may shed light on how stars evolve across the gap.
	
	Finally, while the population of stars within the gap is small, it appears to be nonzero, which opens up the possibility of targeted studies of stars that are currently crossing the gap. Detailed observations of individual stars in the gap or near the lower boundary of the gap may reveal interesting aspects of their activity and the processes that shepherd them across this span of the color-period diagram. 
    
\section{Conclusions}\label{sec:conclusions}

    We have measured precise rotation periods for 8,943 main-sequence K2 stars by Gaussian process regression. We perform MCMC simulations on each light curve to obtain estimates of the GP hyperparameters and their uncertainties. We detect and measure the gap in the rotation period distribution and show that this feature appears in all K2 campaigns and is thus unlikely to result from a peculiarity of the stellar populations observed by Kepler. We review several explanations for the gap and argue that the most likely is that the gap results from stalled spindown on the fast-rotating sequence for low-mass stars, followed by rapid evolution across the gap to the slow-rotating sequence. This evolution may be governed by time-variable core-envelope coupling, which controls the rate of transfer of angular momentum from the core to the surface of the star. 
    
    In the future, TESS observations will provide a large sample of light curves for field stars. We expect that a similar distribution of rotation periods will be observed for this sample. One key observation that TESS may enable is whether or not the gap extends to stars more massive than $\sim 0.8 M_\odot$. If the gap represents the space between two separate stellar populations at different ages, then it should extend to higher mass stars, but if the gap emerges from the physics of core-envelope coupling then we may expect to observe a mass-dependence for the phenomenon. 
    
    Finally, one dimension that has been left out of this work is that of metallicity. \cite{Amard2020} reports a metallicity dependence to stellar rotation in the Kepler which may be detectable in our K2 sample as well. As the inner structure and the evolution of a star is known to be dependent on its chemical composition, this dependence of rotation period on metallicity may help to illuminate the relationship between interior structure and spindown. We leave the task of exploring this relationship to future work. 
    
    By making the full results of our MCMC simulations available to the community we hope to make it possible for other researchers to make different choices about which periods to include and exclude. Machine learning techniques such as convolutional neural networks or random forests may also be useful for identifying rotation signals in EVEREST light curves. Combining these techniques with our period measurements may result in a larger final sample without sacrificing quality. Our sample may also be of use as a training set for machine learning algorithms seeking to identify stellar rotation signals. 

\acknowledgements

This research was supported by the National Aeronautics and Space Administration (NASA) under grant number 80NSSC18K1660 issued through the NNH17ZDA001N Astrophysics Data Analysis Program (ADAP).

TAG was supported by NSF grant AST-1907342

This work was facilitated though the use of the advanced computational, storage, and
networking infrastructure provided by the Hyak supercomputer system at the University of
Washington.

JRAD acknowledges support from the DIRAC Institute in the Department of Astronomy at the University of Washington. The DIRAC Institute is supported through generous gifts from the Charles and Lisa Simonyi Fund for Arts and Sciences, and the Washington Research Foundation.

This paper includes data collected by the Kepler mission and obtained from the MAST data archive at the Space Telescope Science Institute (STScI). Funding for the Kepler mission is provided by the NASA Science Mission Directorate. STScI is operated by the Association of Universities for Research in Astronomy, Inc., under NASA contract NAS 5–26555.

This work has made use of data from the European Space Agency (ESA) mission
{\it Gaia} (\url{https://www.cosmos.esa.int/gaia}), processed by the {\it Gaia}
Data Processing and Analysis Consortium (DPAC,
\url{https://www.cosmos.esa.int/web/gaia/dpac/consortium}). Funding for the DPAC
has been provided by national institutions, in particular the institutions
participating in the {\it Gaia} Multilateral Agreement.

\software{Python, IPython \citep{ipython}, 
NumPy \citep{numpy}, 
Matplotlib \citep{matplotlib}, 
SciPy \citep{scipy}, 
Pandas \citep{pandas}, 
Astropy \citep{astropy}
Exoplanet \citep{exoplanet:exoplanet}
Scikit-image \citep{scikit-image}
}




\bibliography{main}



\end{document}